\documentclass[twocolumn,english]{IEEEtran}
\usepackage[T1]{fontenc}
\usepackage[latin9]{inputenc}
\usepackage{color}
\usepackage{array}
\usepackage{verbatim}
\usepackage{multirow}
\usepackage{amsmath,epsfig}
\usepackage{amssymb}
\usepackage{graphicx}
\usepackage{esint}
\usepackage{epstopdf}

\usepackage{cite}
\ifCLASSINFOpdf
\else
\fi
\usepackage{color}
\usepackage[normalem]{ulem}

\begin{document}

\title{\textit{Dithen}: A Computation-as-a-Service Cloud Platform For Large-Scale Multimedia Processing}

\author{Joseph Doyle,
        Vasileios Giotsas,
        Mohammad Ashraful Anam and~Yiannis~Andreopoulos,~\IEEEmembership{Senior~Member,~IEEE}% <-this % stops a space
\thanks{The authors are with Dithen, 843 Finchley Road, London, NW11 8NA, U.K., {http://www.dithen.com}; email: \textbraceleft j.doyle, v.giotsas, russell.anam, i.andreopoulos\textbraceright@dithen.com. J. Doyle is also with the University of East London, University Way,
London, E16 2RD, U.K.
 M. A. Anam and Y. Andreopoulos are also with the Electronic and Electrical Engineering Department, University College London, Roberts Building, Torrington Place, London, WC1E 7JE, UK. V. Giotsas is also with the Center for Applied Internet Data Analysis, University of California at San Diego, CAIDA UCSD/SDSC, 9500 Gilman Dr., La Jolla, CA, 92093-0505. This work was supported in part by Innovate UK (feasibility project ACAME,\ 131983). \ }% <-this % stops a space
\thanks{}}


\markboth{submitted}%
{Shell \MakeLowercase{\textit{et al.}}: Bare Demo of IEEEtran.cls for Journals}



\maketitle

\begin{abstract}
We present \textit{Dithen}, a novel computation-as-a-service (CaaS)\ cloud platform specifically tailored to the parallel execution of large-scale multimedia tasks. Dithen handles the upload/download of both  multimedia data and executable items, the assignment of compute units to multimedia workloads, and the reactive control of the available compute units to minimize the cloud infrastructure cost under deadline-abiding execution. Dithen combines three key properties: \textit{(i)}\ the reactive assignment of individual multimedia tasks to available computing units according to availability and predetermined time-to-completion constraints; \textit{(ii)} optimal resource estimation based on Kalman-filter estimates; \emph{(iii)} the use of additive increase multiplicative decrease (AIMD) algorithms (famous for being the resource management in the transport control protocol) for the control of the number of units servicing workloads. The deployment of Dithen over Amazon EC2 spot instances is shown to be capable of processing more than 80,000 video transcoding, face detection and image processing tasks (equivalent to the processing of more than 116 GB of compressed data) for less than \$1 in billing cost from EC2. Moreover, the proposed AIMD-based control mechanism, in conjunction with the Kalman estimates,\ is shown to provide for  more than 27\% reduction in EC2 spot instance cost against methods based on reactive resource estimation.
Finally, Dithen is shown to offer a  38\% to 500\%\ reduction of the billing cost  against the current state-of-the-art in CaaS platforms on Amazon EC2 (Amazon Lambda and Amazon Autoscale). A baseline version of Dithen is currently available at {http://www.dithen.com} under the ``AutoScale'' option.  \end{abstract}

\begin{IEEEkeywords}
computation-as-a-service, big data, multimedia computing, cloud computing, Amazon EC2, spot instances.
\end{IEEEkeywords}

% \fi
\IEEEpeerreviewmaketitle

\section{Introduction}
\IEEEPARstart{I}{nfrastructure-as-a-service} (IaaS) providers, such as Amazon Elastic Compute Cloud (EC2),  Google Compute Engine (GCE), IBM Bluemix and Rackspace, now allow for the flexible reservation of \textit{compute units} (CUs) in the cloud (i.e., pre-established sets of processor cores, memory, storage and operating systems), with yearly, daily, hourly or even minute-by-minute billing \cite{song2012optimal,zhang2014dynamic}. This has led to the explosion of \textit{Platform-as-a-Service} (PaaS) \   and \textit{Software-as-a-Service} (SaaS)  offerings \cite{nan2012optimal,hobfeld2012challenges}. Within PaaS systems, the user is able to develop and execute processing tasks on distributed computing environments (e.g., Apache Hadoop/mesos, Google App Engine) on IaaS providers, albeit at the cost of converting the multimedia processing software to code that can be scaled-up by the PaaS infrastructure (for example converting the operations  to  Map and Reduce steps in Hadoop).  In SaaS, a provider licenses a specific set of
applications to customers (e.g., pre-established word processing software, a fixed set of video transcoding or video streaming toolboxes, etc.) either as a service on demand, through a
subscription, or in a pay-as-you-go model \cite{islam2012giving}. In the multimedia systems domain, this provides the opportunity to use transcoding or signal processing algorithms and toolboxes directly \cite{andreopoulos2008incremental,spiliotopoulos2001quantization,andreopoulos2001local,andreopoulos2002new,kontorinis2009statistical,hobfeld2012challenges}, and has led to the development  of related commercial services.

\subsection{From Platform and Software-as-a-Service to Computation-as-a-Service}

This evolution of IaaS, PaaS and SaaS is now beginning to lead to \textit{Computation-as-a-Service} (CaaS) \cite{masiyev2012cloud},  where users can upload multimedia (e.g., image, audio or video) files \textit{and} scripts or binary files prepared in their local environment \cite{andreopoulos2008incremental,andreopoulos2000hybrid,kontorinis2009statistical,hobfeld2012challenges,andreopoulos2003high} in order to be executed in CUs in the cloud directly, i.e., without having to develop and manage any infrastructure or convert their software to a format amenable to distributed computing environments.
CaaS provides a useful compromise between the generality of IaaS and PaaS offerings and the  ease-of-use of SaaS: the end user can deploy and scale \textit{any} desktop multimedia application of their choosing \textit{without} needing to adapt its codebase. This differs from the case of SaaS,  in that the user can simply execute any Matlab, C/C++, Java, OpenCV,  Javascript/Python\ based code and scripts of their local platform on the CaaS\ platform without any modification by following a  simple set of rules. The CaaS platform can then handle the scheduling and parallelization of multiple multimedia workloads without any user intervention via the appropriate reservation (or bidding)\ of resources from IaaS providers, e.g., Amazon EC2 spot instance bidding or GCE\ CU reservation.

\subsection{ Related Work}
Workload scheduling on CaaS systems has some resemblance to well-studied scheduling problems for large computing clusters \cite{Schwarzkopf2013Omega,boutin2014apollo,ousterhout2013sparrow}. However, two major differences between the two domains are that resources of cluster computing systems are persistent and ``prepaid'', i.e.,  the number of CUs does not fluctuate during the execution of a workload and there is no penalty for unused CUs. On the other hand, because CaaS resources are billed according to compute instance reservation, a CaaS provider often initiates or terminates CUs during the execution of a workload in order to minimize the monetary cost incurred, while abiding to the agreed workload completion time. To this end, there have been numerous recent proposals   for cloud resource management. Gandhi \textit{et. al.} propose their own version of Autoscale, which stops servers that have been idle for more than a specified time, while concentrating jobs on less CUs to reduce cost \cite{Gandhi2012autoscale}. Paya \textit{et. al.} expand on this by proposing a system that uses multiple sleep states to improve performance \cite{Paya2015loadbalancing}. Song \textit{et. al.} propose optimal allocation of CUs according to pricing and demand distributions \cite{song2012optimal}. Ranjan \textit{et. al.} investigate architectural elements of content-delivery networks with cloud-computing support \cite{ranjan2013mediawise}. Jung \textit{et. al.} propose multi-user workload scheduling on various CUs based on genetic algorithms \cite{jung2014estimation}. 
Beyond resource allocation and scheduling, a major challenge in CaaS frameworks is the varying delay in the completion of various multimedia processing workloads \cite{hobfeld2012challenges,gulisano2012streamcloud}. The processing delay primarily depends on: the workload specifics, the CU reservation mechanism employed, and the transport-layer jitter (if data is continuously transported to/from users and cloud providers) \cite{islam2012giving}. This is the primary reason why all real-world CaaS platforms only provide ``best effort" service level agreements (SLAs) for large workload execution without considering a predetermined time-to-completion (TTC)\ estimate. Recent research work on this front proposes the use of particle swarm optimisation to derive viable schedules  \cite{Rodriguez2014deadlinescheduling} and the use of the earliest-deadline first algorithm \cite{Mao2011deadlinescheduling}. While all such proposals are effective in their resource provisioning for TTC-abiding execution, they assume that the system has accurate estimates of the computation required to complete each workload. However, this is unlikely to be the case in practice, particularly at the start of a workload's execution. Therefore, our proposal considers the realistic scenario where no estimates for the computational requirements are available at the start of each workload's execution; i.e., in conjunction with resource provisioning, our framework performs an adaptive resource estimation \textit{during} the execution of each workload.

Finally, the first commercial CaaS\ offerings are now  beginning to emerge. The key representatives are: \textit{(i)}\ the recently-announced AWS Lambda service, where users can submit individual Javascript items and be billed at a fixed rate per 100ms of Lambda service usage under a best-effort SLA;  \textit{(ii)}\ PiCloud,\ a service for flexible scheduling of batch processing tasks via a terminal command line interface; \textit{(iii)}\ Parse, a software development environment for Javascript execution on cloud-computing infrastructures; and \textit{(iv)} Amazon EC2 Autoscale, a service that automatically scales application deployment over Amazon EC2 according to processor and network utilization constraints. In all these deployments, the comparative metric for workload analysis is the required processing time in terms of the number of seconds a single core was occupied until the workload is successfully completed. We therefore quantify the resource reservation in the IaaS provider via \textit{compute-unit seconds} (CUSs), i.e., the product of the total cores used  with the time they were reserved for, since charges will be applied for them from the IaaS provider regardless of whether the CaaS system actually used them to their full capacity or not.

\subsection{ Contribution}

While the current research and commercial efforts in CaaS frameworks are a promising start, they do not consider  the reactive estimation of the required CUSs to process submitted workloads, or assume that the CUS metric per workload is known \cite{hobfeld2012challenges,Rodriguez2014deadlinescheduling,Mao2011deadlinescheduling}. In addition, current CaaS\ frameworks do not consider on-demand CU provisioning (e.g., EC2 spot instances or GCE\ CUs with minute-level increments) under TTC\ constraints, where it is imperative to control \textit{both} the allocation and termination of new instances in order to reduce the infrastructure cost while providing for TTC-abiding execution. Finally, at the moment there are very limited options for  CaaS frameworks to develop and benchmark multimedia cloud-computing services and the multimedia systems community would benefit from new efforts on this front. \ 

In this light, we present \textit{Dithen}, a new cloud computing service that scales small and medium-level execution of data processing workloads to big data under TTC constraints. For example, algorithms for video transcoding, image classification, object recognition, etc., that run on small volumes of input images/videos on a desktop computing system can be directly scaled-up via Dithen (i.e., without any code modifications)  to operate on big datasets comprising millions of input images and videos, with \textit{a-priori} established completion times. Dithen meets the requirements for such large-scale data-intensive processing by  combining the following novel aspects:\  

\begin{enumerate}
\item 
It supports the direct upload and execution of bash, Python, Java, Javascript and Matlab scripts, as well as the execution of statically-built binaries for 32-bit or 64-bit Ubuntu Linux or Microsoft Windows on any number of EC2 spot instances.  

\item
Each submitted workload is separated into individually-executable tasks, which are then allocated to available CUs with proportionally-fair scheduling in order to: \textit{(i)} maximize the available CU\ utilization and \textit{(ii)} abide by the confirmed TTC value for the workload.
The fine-grain partitioning of each workload into tasks allows for each user to check that the output results are being produced correctly by Dithen during execution and cancel the workload execution if otherwise.

\item
Estimates of the required CUSs until the completion of each task type in each workload are derived based on Kalman-filter estimators, which are shown to significantly outperform other \textit{ad-hoc} estimators.  

\item 
Based on the estimation of the required CUSs, Dithen uses the Additive Increase Multiplicative Decrease (AIMD)\ algorithm \cite{shorten2006positive}    for the allocation or termination of CUs according to the expected workload.
While AIMD is a well-known control mechanism for network resource utilization, e.g., within the transport control protocol (TCP), to the best of our knowledge, this is the first time it is proposed for CaaS provisioning.

\end{enumerate}\

 Finally, beyond describing Dithen, we also provide free access to it\footnote{each new user account gets an amount of free credit to spend on the service} at {http://www.dithen.com} under the ``AutoScale'' option of the main file manager service.

The remainder of this paper is organized as follows.   Section \ref{Sec2} presents the architecture of Dithen. Sections \ref{Sec3} and \ref{Sec4}  present the key elements of the proposed CUS estimation  and AIMD framework, while Section \ref{Sec5} presents experimental results and comparisons of different CU\ allocation strategies for Amazon EC2 spot instances. Finally, Section \ref{Sec6}\ presents some concluding remarks.

\section{Anatomy of the Dithen Architecture }\label{Sec2}

\begin{figure}[t]
\includegraphics[scale=0.22]{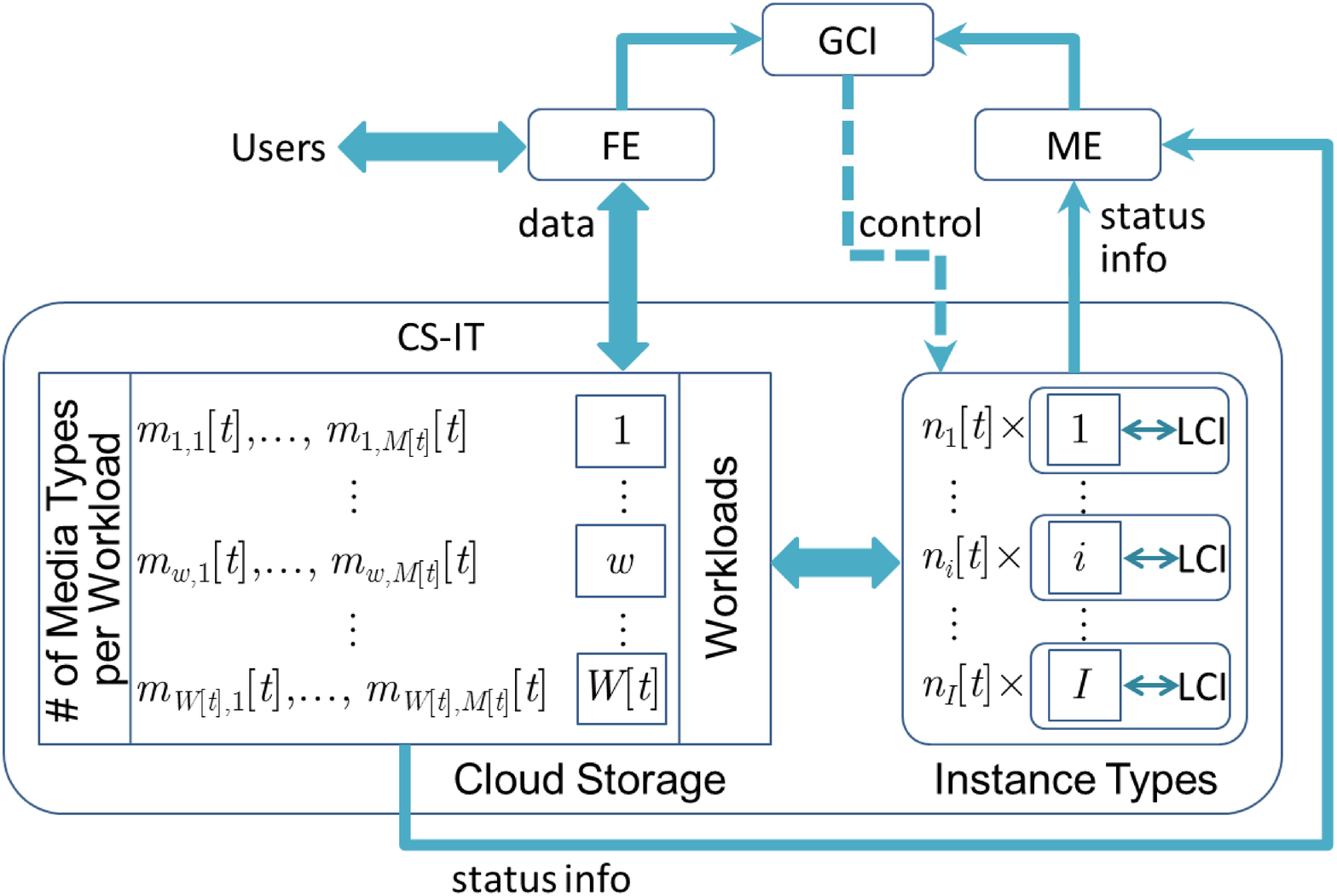}
\centering

\caption{Dithen architecture. At each monitoring instant $t$, each workload $w$ ($1 \leq w \leq W[t]$)\ contains several media types. In addition, $n_i[t]$ spot instances of type $i$ ($1\leq i \leq I$) are reserved and can be used to process workloads.  }
\label{fig:FigDithen}
\end{figure}

The architecture of Dithen is pictorially illustrated in Fig. \ref{fig:FigDithen}. It comprises five elements: the Front End (FE), the Cloud Storage and Instance Types (CS-IT), the Monitoring Element (ME),  and the Local and Global Controller Instances (LCI and GCI). Their functionality is  detailed in the following subsections.
To aid the exposition, Table \ref{tab:Notation-table} summarizes the  nomenclature and notational conventions used.

\begin{table}
\noindent \centering{}\protect\caption{\label{tab:Notation-table}Nomenclature and Notational Conventions.}
\begin{tabular}[t]{>{\centering}p{0.2\columnwidth}>{\raggedright}p{0.7\columnwidth}}
\hline 
\noalign{\vskip\doublerulesep}
\textbf{Key Concept} & \centering{}\textbf{Definition}\tabularnewline[\doublerulesep]
\hline 
\noalign{\vskip\doublerulesep}
\centering{}$t$ & monitoring time instant \tabularnewline[\doublerulesep]
\noalign{\vskip\doublerulesep}
\centering{}$W[t]$ & total workloads in Dithen at time instant $t$\tabularnewline[\doublerulesep]
\noalign{\vskip\doublerulesep}
\centering{}$M[t]$ & total media types at time \(t\)\tabularnewline[\doublerulesep]
\noalign{\vskip\doublerulesep}
\centering{}$m_{w,k}[t]$ & remaining media items of type $k$ to be processed within workload $w$ at time \(t\) $( 1 \leq k \leq M[t]$, $1 \leq w \leq W[t] )$ \tabularnewline[\doublerulesep]
\noalign{\vskip\doublerulesep}
\centering{}$I$ & total \textit{types} of instances in the cloud infrastructure\tabularnewline[\doublerulesep]
\noalign{\vskip\doublerulesep}
\centering{} $p_i$\   & compute units (CUs), i.e., processor cores, available within instance type \(i,1\leq i \leq I\) \tabularnewline[\doublerulesep]
\noalign{\vskip\doublerulesep}
\centering{}$n_i[t],N_\text{tot}$ & \textit{number} of instances of type \(i\) (\(1\leq i \leq I\)) reserved at time \(t\), total \textit{number} of CUs in Dithen \tabularnewline[\doublerulesep]
\noalign{\vskip\doublerulesep}
\centering{}$a_{i,j}[t]$\   & remaining time for the \(j\)th instance of type \(i\) before additional billing is incurred by the cloud provider  \tabularnewline[\doublerulesep]
\noalign{\vskip\doublerulesep}
\centering{}$c_{\text{tot}}[t],c_{\text{min}},c_{\text{max}}$\   & total compute-unit-seconds (CUSs) available in Dithen, and lower/upper limits for CUSs in Dithen  \tabularnewline[\doublerulesep]
\noalign{\vskip\doublerulesep}
$d_{w}[t]$ & time-to-completion (TTC) for workload \(w\) at time \(t\)\tabularnewline[\doublerulesep]
\noalign{\vskip\doublerulesep}
$\hat{b}_{w,k}[t]$ & CUS estimate to process a media item of type \(k\) of workload \(w\)\ \tabularnewline[\doublerulesep]
\noalign{\vskip\doublerulesep}
$r_{w}[t]$ & required CUSs for the completion of workload \(w\) \tabularnewline[\doublerulesep]
\noalign{\vskip\doublerulesep}
$s_{w}[t]$ & service rate, i.e., CUs allocated for workload \(w\) \tabularnewline[\doublerulesep]
\noalign{\vskip\doublerulesep}
$z_{w,k}[t],v_{w,k}[t]$ & CUS process and measurement noise instantiations of media type \(k\) of workload \(w\) \tabularnewline[\doublerulesep]
\noalign{\vskip\doublerulesep}
$\alpha,\beta$ & additive increase and multiplicative decrease parameters of AIMD \tabularnewline[\doublerulesep]
\hline 
\noalign{\vskip\doublerulesep}
\textbf{Notation} & \centering{}\textbf{Explanation}\tabularnewline[\doublerulesep]
\noalign{\vskip\doublerulesep}
\hline 
\noalign{\vskip\doublerulesep}
uppercase Roman letters & random variables\tabularnewline[\doublerulesep]
\noalign{\vskip\doublerulesep}
lowercase Greek letters & moments of probability distributions, stochastic parameters of Kalman filters, or AIMD and ARMA\ parameters\tabularnewline[\doublerulesep]
\noalign{\vskip\doublerulesep}
$\tilde b$ & measurement of quantity \(b\)\tabularnewline[\doublerulesep]
\noalign{\vskip\doublerulesep}
$\hat b$ & estimation of quantity \(b\)\tabularnewline[\doublerulesep]
\noalign{\vskip\doublerulesep}
\hline 
\noalign{\vskip\doublerulesep}
\end{tabular}
\end{table}

\begin{figure}[t]
\includegraphics[scale=0.22]{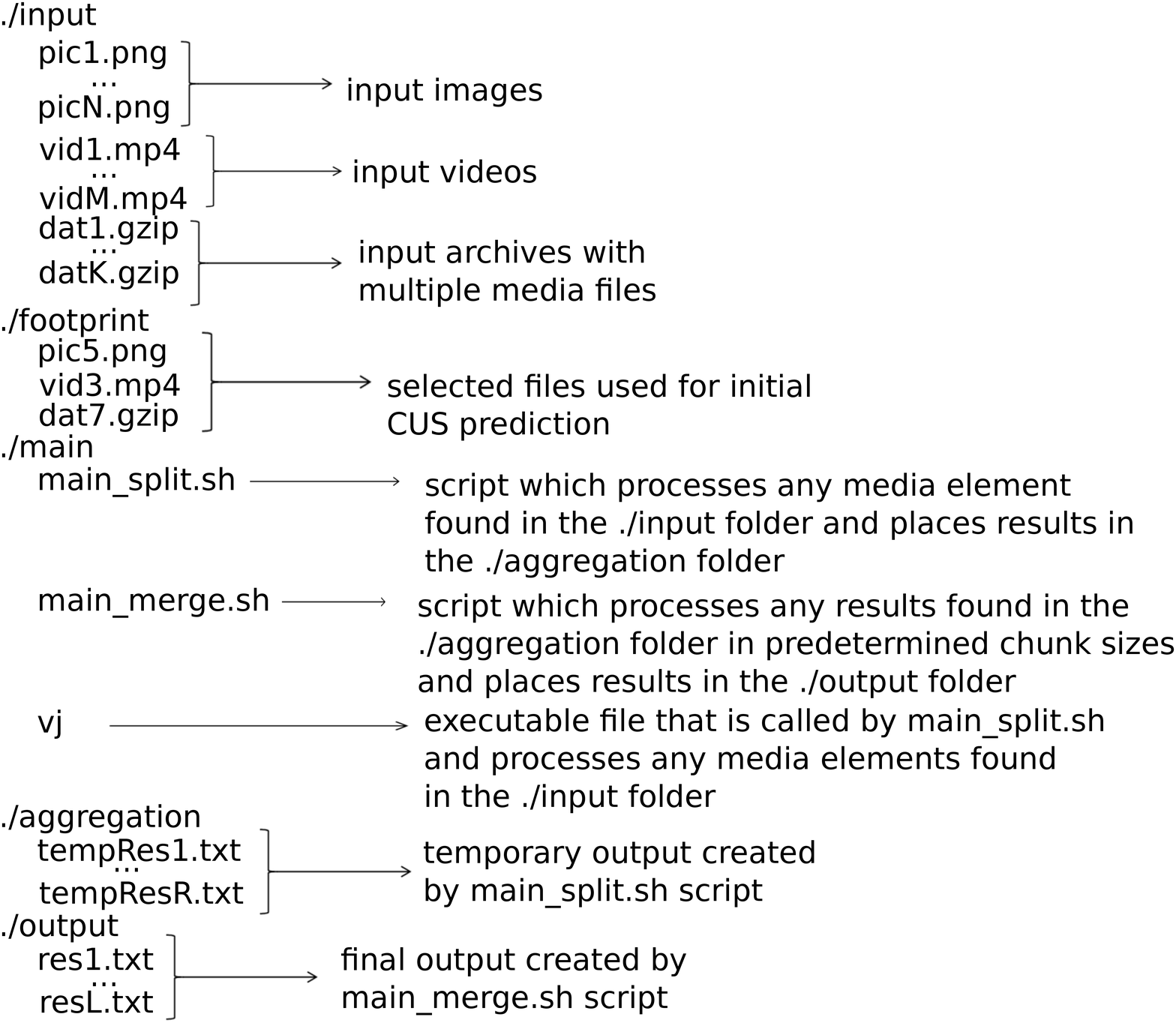}
\centering

\caption {{Example workload structure.}}
\label{fig:FigWorkloadStructure}
\end{figure}

\subsection{ Overview of the Operation of Dithen}
In order to illustrate the roles of the different components of our system, we first present an overview of how Dithen processes a workload. A user submits a workload via the FE (Section \ref{sec:FE}) and can request a particular TTC value for this workload, or a TTC is allocated by Dithen.  {As shown in Fig. \ref{fig:FigWorkloadStructure}}, the workload may comprise multiple input media files (e.g., JPEG images, MP4 video files, etc.), as well as scripts and executable files to process the inputs. {An executable file can be a Linux or Windows binary and the script can be a Linux shell script, Windows batch file or Matlab/Octave  script file.} Dithen typically handles workloads where each input is processed independently from the other inputs. This is the norm for large-scale multimedia cloud computing where a user typically wants to carry out a certain task (e.g., face recognition, transcoding, etc.) on a large cache of input images or videos. {Nevertheless, as explained in later sections (and as shown in Fig. \ref{fig:FigWorkloadStructure}) it is also possible to: \textit{(i)}  package multiple inputs together in a single input archive (e.g., gzip file, the contents of which are automatically extracted by Dithen in the utilized CU prior to processing) for concurrent processing of batches of input media, \textit{(ii)}\  execute ``Split-Merge'' tasks akin to MapReduce code, i.e., code that performs parallel processing of inputs (Split step)\ followed by aggregation of multiple outputs in order to produce the final results (Merge step).}

Once  the GCI detects that a new workload has been added, it assigns a small percentage of the inputs of the workload (e.g., 5\% of the submitted inputs) to LCIs in a ``footprinting" stage.  The compute instances of the LCIs execute the submitted code on their assigned inputs and provide the corresponding execution times to the ME (Section \ref{sec:ME}) and, via that, to the GCI. These measurements and the logs of the execution status (e.g., 0 for normal and -1 for abnormal termination) are used by GCI to: \textit{(i)} confirm that the workload processing is carried out without errors or crashes in the submitted code; \textit{(ii)} derive an initial Kalman-based estimate of the required CUSs to complete this workload (Section \ref{sec:LCI_GCI}-1). This estimate is used to confirm that the requested workload TTC  is achievable by Dithen, or else adjust the confirmed TTC accordingly  (Section \ref{sec:LCI_GCI}-2). The GCI continues to derive CUS estimates per workload in order to: \textit{(i)} assign a service rate per workload, according to which all LCIs can process workload tasks via their corresponding instances (Section \ref{Sec3}); \textit{(ii)}  determine if  the number of CUs should be scaled up or down (via the proposed AIMD algorithm of Section \ref{Sec4}) so that all confirmed TTCs are met without excessive billing from the cloud provider. Finally, the results produced by all CUs are uploaded to Amazon Simple Storage Service (S3, see Section \ref{sec:CS-IT}), where they can be viewed by the user through the Dithen FE.

\subsection{Front End and Workload Processing Modes}\label{sec:FE}

\begin{figure}[t]
\includegraphics[scale=0.16]{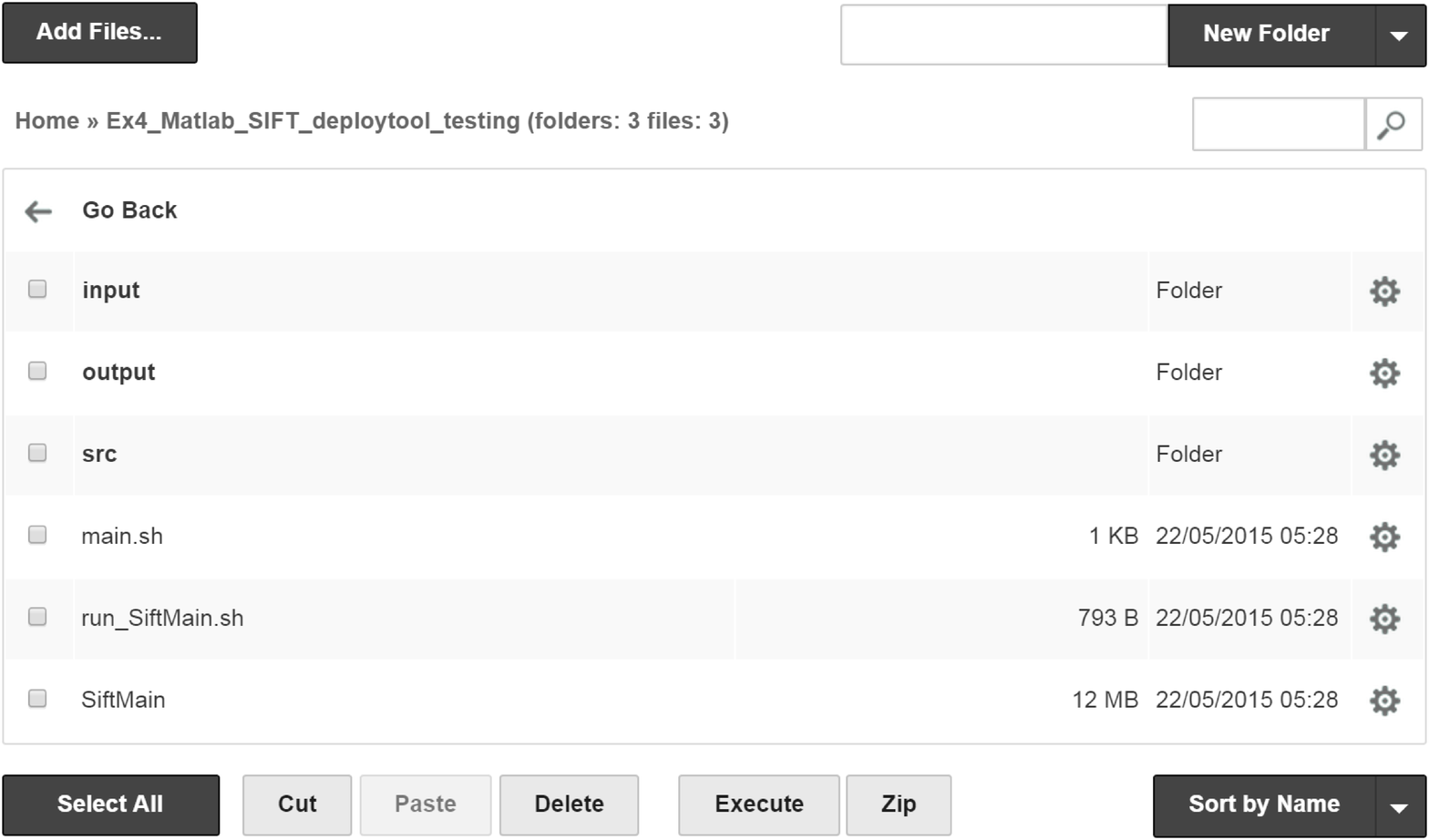}
\begin{footnotesize}\begin{center}(a)\end{center}\end{footnotesize}

\includegraphics[scale=0.16]{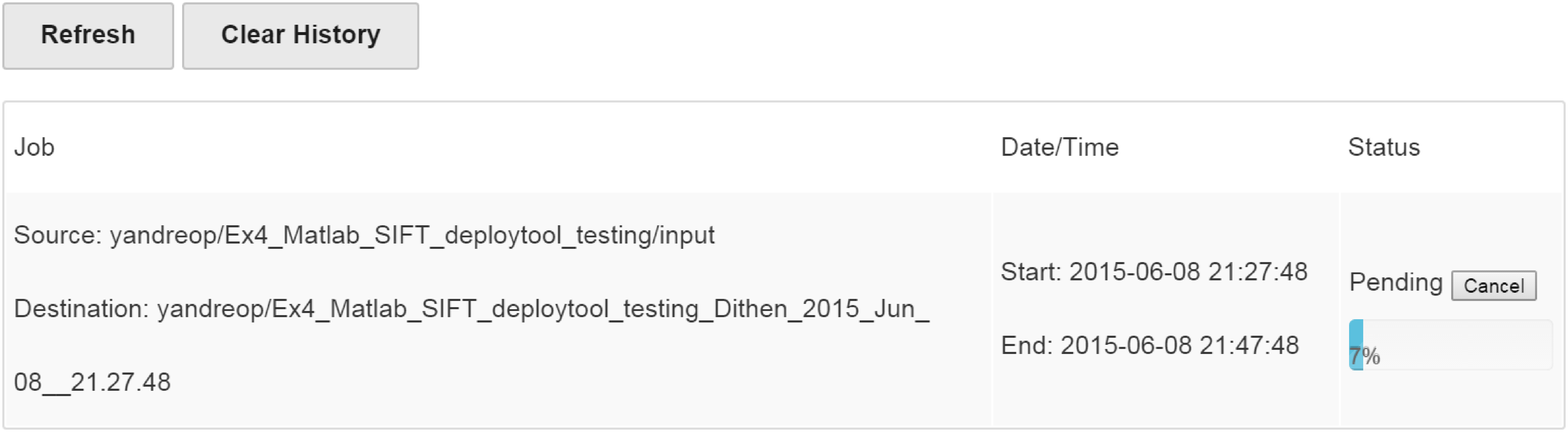}
\begin{footnotesize}(b)\end{footnotesize}
\centering

\caption{(a) Example of front-end contents for a user workload executing the SIFT\ feature extraction on a large image dataset. The input image dataset is contained in the \texttt{./input} folder and the output results are produced in the \texttt{./output} folder. (b) Example of the workload execution indicating the source and destination folders and the confirmed TTC.   \ }
\label{fig:FigFrontEnd}
\end{figure}

The FE of Dithen provides for workload uploading, launching, monitoring of execution, basic text file and image viewing and editing functionalities (e.g., for log file or launch script viewing and editing), and downloading of the results of individual tasks as they are being produced by CUs.   
 Users can utilize the FE\ to cancel pending workloads if the results are deemed to be unsatisfactory or incorrectly executed (e.g., due to unsupported runtime components, or crashes/errors  in the user's executed code). The simplicity of the FE of Dithen is evident from Fig. \ref{fig:FigFrontEnd}: the buttons allow for a point-and-click interface via a web browser. A basic mobile FE interface is also available.
 
 \textbf{1. Code \&\ data elements and baseline workload processing mode:}\ In the basic mode of operation, Dithen assumes that the user provides a ``main" bash/batch script (i.e., \texttt{main.sh} in Linux or \texttt{main.bat} in Microsoft Windows) for each application, which invokes all the required Matlab,  Python, Java/Javascript code, or application binaries. All provided multimedia elements must be available in the folder \texttt{./input/} within each application. The results are produced in the locations created and specified within the user's own code (typically in application folders such as \texttt{./output/} or \texttt{./results/}, etc.). The only constraint imposed by Dithen is that results must not be created within the input folder, which should be reserved solely for input files. An example of a typical input/output application structure is given in Fig. \ref{fig:FigFrontEnd}(a) and an example of the FE for the job monitoring is given in Fig. \ref{fig:FigFrontEnd}(b). 

Each new user account created by the FE\ includes by default (at least)\ four example image processing applications that illustrate how to use the service:\ \textit{(i)} \texttt{Ex1\_face\_detection}---automated detection of human faces within individual images using a stand-alone C++ implementation of the Viola-Jones algorithm \cite{viola2004robust}; \textit{(ii)} \texttt{Ex2\_template\_match}---template matching between input images with an existing library of template images using the ImageMagick \texttt{compare} tool \cite{imagemagick}; \textit{(iii)} \texttt{Ex3\_image\_merge}---merging of groups of input JPEG\ files into a single animated GIF\ file using the ImageMagick \texttt{convert} tool \cite{imagemagick}; \textit{(iv)} \texttt{Ex4\_Matlab\_SIFT}---image salient point detection and description using the SIFT algorithm \cite{lowe2004sift} via a Matlab implementation compiled to stand-alone binary with the Mathworks \texttt{deploytool}. 

In all cases, the provided code utilizes each input media file independently and populates the output folder(s) with the results of the executed algorithm(s). Each such execution comprises  a \textit{media processing task}. The entire volume of independently-processed inputs, along with the user's application code, comprises a \textit{workload}. If a number of media inputs (e.g., images, videos and/or audio files) must be processed together by the provided application code, they must be packaged into a single file container\footnote{This is illustrated in example   \texttt{Ex3\_image\_merge}.} (e.g., TAR, ZIP or RAR).  Dithen will automatically extract all such compressed format containers prior to calling the main script. 

\textbf{2. Advanced workload processing mode:} To allow for more complex interactions between inputs and subsequent results, such as split and merge tasks that are similar to MapReduce code \cite{Dean2008MapReduce}, Dithen provides an advanced workload processing mode,  specifically designed to allow for the execution of Split-Merge tasks, e.g., the parallel execution of visual feature extraction from input JPEG images followed by aggregation of the produced feature points into a single feature matrix of reduced dimensions \cite{abbas2015vectors,chadha2015region}. {This mode is triggered by the user uploading two ``main'' scripts, called \texttt{main\_split.sh} and \texttt{main\_merge.sh} (see also Fig. \ref{fig:FigWorkloadStructure}). When these scripts are found in the application folder, the first one is executed as in the basic mode of operation described previously but, instead of returning the results to the FE, it uploads them to a specially designated  ``aggregation'' spot instance in Amazon EC2 that runs the second (i.e., ``Merge'') script. In this way, the latter script can  invoke any aggregation code and the final results are provided to the FE. For example, within image retrieval or face recognition applications \cite{yang2004two,abbas2015vectors,chadha2015region}, the user can parallelize the computation of a very large number of image covariance matrices or vectors of local features via the ``Split'' script and then perform a large singular value decomposition (SVD) of the results (once they become available) by having the ``Merge'' script periodically poll for the full set of results of the Split step and invoke the SVD code on them. The data and executable items corresponding to such an example are  illustrated in Fig. \ref{fig:FigWorkloadStructure}: the Split step is launched simultaneously by multiple instances running the \texttt{main\_split.sh} script (each containing subsets of inputs) and each instance calculates one or more of the \texttt{tempRes1.txt...tempResR.txt} results and places them in the \texttt{/aggregation} folder of a specially-designated ``Merge'' instance. This instance is running the \texttt{main\_merge.sh} script periodically in order to poll the \texttt{/aggregation} folder and, once sufficient outputs are detected, produces each of the \texttt{res1.txt...resL.txt} files based on groups of such outputs. The rule of how many (and which)\ outputs to poll for, as well as the polling frequency (e.g., once per minute), is set by the user within the \texttt{main\_merge.sh} script.}

\subsection{Cloud Storage and Instance Types}\label{sec:CS-IT}

The CS-IT deployment depends on the possibilities available by the IaaS provider. In this paper, we evaluate Dithen with Amazon EC2 spot instances and Amazon S3 storage. We opt for EC2 spot instances as they provide for a wide variety of available configurations and for flexible billing based on hourly reservations. Future evaluations can incorporate other IaaS providers, like GCE, IBM Bluemix and Rackspace. 

Dithen uses Ubuntu Linux and MS\ Windows spot instances that have been setup with bash shell support (or batch script file support for MS Windows), Python, Java, Javascript, Matlab (via the Mathworks Matlab Compiler Runtime), and OpenCV and ImageMagick library support. As shown in Fig. \ref{fig:FigDithen}, depending on the type of scripts and executables submitted by the user, either of these instances can be spawned into any number of  spot instances of type \(i\), out of \(I\) total instance types. This is performed by bidding for an appropriate spot instance type  in Amazon's EC2 launch process  (out of more than twenty types available) and specifying the reserved Amazon machine image id to be spawned.  We denote the number of CUs per instance type by $p_i$, $1\leq i \leq I$. Moreover, at the \(t\)th time instant, the Dithen architecture contains $n_i[t]$ total instances of type $i$. Finally, for the $j$th instance out of the  $n_i[t]$ ones, the remaining time until the next billing increment (e.g., until the time when the IaaS provider will bill for the next hour in the corresponding spot instance) is denoted by $a_{i,j}[t]$ (in seconds). Spot instances are requested using the \texttt{requestSpotInstance()} function which is part of the EC2 class of the AWS SDK. Instances are terminated using the \texttt{terminateInstances()} of the same class. Finally, the number of active spot instances is monitored using the \texttt{describeInstances()} function of the same class.

\subsection{Monitoring Element}\label{sec:ME}

In order to observe the CU utilization, Dithen includes a monitoring element that measures processor utilization within each spot instance via the \texttt{mpstat} Linux command (or \texttt{wmic cpu} in MS Windows). Monitoring and reactive control of the execution of workloads take place at discrete ``monitoring'' time  instants, typically every 1--5 minutes.   The ensemble of all the currently executing workloads within Dithen at the $t$th monitoring instant includes $M[t]$ different media types (e.g., images, audio, video files, or container files comprising composite media and data types). The ME keeps track of a number of operational parameters described below (and summarized in Table \ref{tab:Notation-table}). 

At every monitoring instant $t$, and within each workload $w$ ($1\leq w \leq W[t]$), the ME\ keeps track of the number of remaining elements to be processed, $m_{w,k}[t]$,
as well as the estimated CUSs required to complete the processing of each media type $k$ with the workload,  $\hat{b}_{w,k}[t]$. The values of $m_{w,k}[t]$ are determined using an SQL database that records which tasks have been processed and the \texttt{getIterator(`ListObjects')} function of the S3 class of the AWS SDK. The estimates $\hat{b}_{w,k}[t]$ are derived based on the estimation process described previously. Typically, the SLA for each workload includes execution within a predetermined TTC value,  $d_w[t]$, which is confirmed after an initial CUS estimate is available for the workload. To this end, the ME continuously keeps track of the required CUSs to complete each workload \(w\), $r_w[t]$, which can be estimated by:  
\begin{equation}
r_w[t]=\sum_{k=1}^{M[t]}m_{w,k}[t]\hat{b}_{w,k}[t].
\label{eq:r_w}
\end{equation}
Finally, the ME\ keeps track of the total number of active CUs in Dithen by:
\begin{equation}
N_{\text{tot}}[t]=\sum_{i=1}^{I} p_{i}n_i[t] ,
\label{eq:N_tot}
\end{equation}   
as well as the total compute-unit seconds billed (i.e., already paid to the IaaS provider and available to use) within the   Dithen architecture at any given instant \(t\):
\begin{equation}
c_\text{tot}[t]=\sum_{i=1}^I\sum_{n=1}^{n_i[t]}p_{i}a_{i,n}[t].
\label{eq:c_tot}
\end{equation}
Effectively, $c_\text{tot}[t]$ and $N_\text{tot}[t]$ represent a ``snapshot'' of the compute resources in Dithen at the $t$th time instant, as they comprise the  available CUSs and CUs under the already-billed EC2 instances.

\subsection{Local and Global Controller Instances}\label{sec:LCI_GCI}

The main tasks at every monitoring time instant $t$ are: \textit{(i)} to ensure that each workload \(w\) is executed within its confirmed TTC, $d_w[t]$, and \textit{(ii)} to match $c_\text{tot}[t]$ to $\sum_{w=1}^{W[t]} r_w[t]$. Both must be met with the minimum billing from the IaaS. These two tasks are accomplished by the LCI and GCI components of Fig. \ref{fig:FigDithen}, respectively.
Towards this end, the
most crucial aspects are: \textit{(i)} defining reliable CUS estimates,  $\hat{b}_{w,k}[t]$, for each media type $k$ within each workload \(w\), \textit{(ii)} confirming the feasibility of each workload's TTC value and selecting the appropriate service rate (i.e., selecting how many CUSs should be allocated to each workload's tasks),   and \textit{(iii)} devising and executing an  algorithm to initialize or terminate CUs according to the demand volume. The first two items allow for \textit{microscale} (i.e., local) control of Dithen, and they are discussed in parts 3 and 4 of this section, as well as in Sections \ref{Sec3} and \ref{Sec4}.
 The last item allows for \textit{macroscale} (i.e., global) control of the workload execution within Dithen;  solutions for these aspects are analyzed in the first two parts of this subsection.

\textbf{1. Task allocation and tracker operation via the GCI:} To achieve each workload's TTC, the GCI divides the workload into chunks and sends these chunks to be processed by the spot instances of LCIs. Specifically, when a workload is submitted, the GCI examines the \texttt{.\textbackslash input} folder to determine how many individually-executable tasks are present in the workload. Once this has been determined, the GCI executes a small number of the tasks in a ``footprinting'' stage. The goal of the ``footprinting'' stage is to determine:\ \textit{(i)} an initial workload CUS estimate per input type; \textit{(ii)} what chunk size to use (i.e., how many inputs to group together for execution by a single spot instance) such that the chunk processing time is comparable to the time interval between monitoring instances (described in Section \ref{sec:ME}). Importantly, while the initial CUS estimate forms the basis for resource estimation in Dithen, it is often inaccurate because it is difficult to select a representative subset of the tasks when the execution time of these tasks is data dependent. For example, in many  face detection  \cite{viola2004robust} or transcoding workloads \cite{garcia2010study}, the estimate that uses only ``footprinting" data can be 50\% higher than the final measured value because of the data dependency of these tasks. Another reason for such inaccuracy is that, when considering small subsets of tasks in some workloads (like Matlab-based applications), the CUS\ estimation is significantly offset by the disproportional amount of time needed to set up the execution environment (a.k.a. ``deadband'' time) in comparison to the actual code execution.  Long deadband times in tasks mandate the grouping of several tasks into large chunks. Once the chunk size has been determined, the GCI connects to the LCIs via the XMLRPC protocol and instructs the LCI to execute the tasks in the chunk. The LCI writes entries to a MySQL database detailing the status of each task as it processes them, as well as execution time measurements once the task is completed. These are used in the Kalman estimation process of Subsection \ref{sec:LCI_GCI}-3. The GCI uses this database to determine which task should be placed in a chunk for an LCI in a  manner analogous to a BitTorrent tracker \cite{Pouwelse2005tracker}: the controller connects to the database to determine which tasks appear as: ``pending'', ``processing'' and ``completed'' and, based on the workload service rates (Section \ref{Sec3}), carries out the chunk allocation to available LCIs. This decoupling between the database writing by the LCIs and the database reading by the GCI prevents bottlenecks and minimizes the network traffic between the GCI and LCIs. A workload is marked as ``completed'' once the GCI\ detects that all tasks in the workload have been completed.

\textbf{2. Spot instance initiation and termination via the GCI:} A direct way to implement the scaling of the required instances is for the global controller instance to constantly match the total CUs billed in Dithen [$c_\text{tot}[t]$ of \eqref{eq:c_tot}] to the total CUs required by all workloads ($\sum_{w=1}^{W[t]} r_w[t]$) at each time instant $t$  by initializing or terminating spot instances (a.k.a. ``reactive''\ control \cite{anderson2012optimal}). However, such an approach is not optimal for the following reasons: \textit{(i)}  $\sum_{w=1}^{W[t]} r_w[t]$   depends on the estimated CUSs required to complete the processing of each media type $k$ within each workload $w$; these estimations will not be accurate for all time instants and media types, and this will lead to unnecessary expenditure to initiate and pay for instances that may never be used due to estimation mismatch; \textit{(ii)}  due to the CU billing for large time intervals (e.g., Amazon EC2 spot instances are billed for one hour and GCE instances are billed in 10-minute slots), as well as the associated delay in initialization or termination of instances (in the order of minutes), rapid fluctuations in $\sum_{w=1}^{W[t]} r_w[t]$   (e.g., due to new workloads or workload cancellations by users)\ will cause bursts of initiation or termination requests  and substantially increased ``dead''\ time, which will be billed by the IaaS; \textit{(iii)} without a control mechanism in place to absorb rapid fluctuations in demand, a flurry of spot instance requests may inadvertently cause unwanted spikes in spot instance pricing \cite{song2012optimal}. In the next two sections, we present our GCI\ proposal for best-effort TTC-abiding execution   that ensures proportional fairness amongst all submitted workloads in Dithen.

\textbf{3. Reliable CUS estimates for media types via Kalman-filter realization:  }Due to the aforementioned inaccuracy of the CUS estimation based on the ``footprinting'' process, we propose the use of an adaptive CUS estimator that runs continuously during the execution of each workload.  In our proposal, each LCI measures the average  CUSs, \(\tilde{b}_{w,k}\), required for each media type \(k\) of each workload \(w\) running on its instance types, by measuring the time to complete tasks between the previous and the current monitoring instance (\(t-1\) and \(t\)) and refining the measurement. We model this measurement operation mathematically by:
 
\begin{equation}
\forall w,k,t: \tilde{b}_{w,k}[t]=\hat{b}_{w,k} [t]+ v_{w,k}[t],
\label{eq:kalman_measurement}
\end{equation}
where   \(v_{w,k}[t]\) is the measurement noise that deviates  $\tilde{b}_{w,k}[t]$ from the ideal  CUS estimate $\hat{b}_{w,k} [t]$  at time instant \(t\). We assume that    \(v_{w,k}[t]\)  can be modeled by  independent, identically distributed (i.i.d.), zero-mean Gaussian random variables,  i.e., $\forall w,k:\mathrm{V}_{w,k} \sim \mathcal{N}\left( 0,\sigma_{v}^2 \right)$.

We express the LCI estimation of the required CUSs for each workload and task type at time \(t\) by: 
\begin{equation}
\forall w,k,t: \hat{b}_{w,k}[t]=\hat{b}_{w,k} [t-1]+ z_{w,k}[t],
\label{eq:kalman_estimation}
\end{equation}
with $z_{w,k}[t]$ the process noise \cite{anderson2012optimal}, expressing variability in the execution time of each task type in each workload across time. We assume that   $\forall w,k:z_{w,k}[t]$  can be modelled by  i.i.d., zero-mean Gaussian random variables, i.e., $\forall w,k:\mathrm{Z}_{w,k} \sim \mathcal{N}(0,\sigma_{z}^2)$. Given  \eqref{eq:kalman_measurement} and \eqref{eq:kalman_estimation} and the fact that all noise terms are i.i.d., the noise variances are: $E \{ \mathrm{V}_{w,k}^2 \}  = \sigma_v^2 $, $E\{ \mathrm{Z}_{w,k}^2  \} = \sigma_z^2 $ and the noise covariance is $E\{ \mathrm{V}_{w,k} \mathrm{Z}_{w,k} \}  = 0$. 

For the measurement and estimation model of   \eqref{eq:kalman_measurement} and \eqref{eq:kalman_estimation}, the optimal estimator for  $\hat{b}_{w,k} [t]$   is known to be the Kalman filter \cite{anderson2012optimal}, which provides for the following two time-update equations for our case ($\forall w,k,t$):

\begin{equation}
 \;\pi^-_{w,k}[t]=\pi_{w,k}[t-1]+\sigma_{z}^2, 
\label{eq:kalman_process_update}
\end{equation}

\begin{equation}
\kappa_{w,k}[t]= \frac{\pi^{-}_{w,k}[t]}{\pi^{-}_{w,k}[t]+\sigma_{v}^2}, 
\label{eq:kalman_coeff_update}
\end{equation} 
where $\pi^{-}$ represents the initial update of the process covariance noise $\pi$, and  $\kappa_{w,k}[t]$ is the Kalman gain of the \(k\)th task type of the $w$th workload at time instant $t$. Based on \eqref{eq:kalman_process_update} and \eqref{eq:kalman_coeff_update}, the estimation of  $\hat{b}_{w,k} [t]$   and the noise covariance update can be written as ($\forall w,k,t$):

\begin{equation}
 \hat{b}_{w,k}[t]= \hat{b}_{w,k} [t-1] + \kappa_{w,k}[t] \left( \tilde{b}_{w,k}[t-1] - \hat{b}_{w,k} [t-1] \right), 
\label{eq:kalman_measurement_update}
\end{equation}

\begin{equation}
 \pi_{w,k}[t]=\left ( 1-\kappa_{w,k}[t] \right)\pi^-_{w,k}[t]. 
\label{eq:kalman_process_update2}
\end{equation} 
   
   \textit{Initialization of proposed CUS estimator per workload and task type}:\ For $t=0$ and $\forall w,k$, the GCI\ initializes each Kalman-filter estimator with $\tilde{b}_{w,k}[0],$ established via  the initial ``footprinting'' measurement per workload and input type, and sets: $\hat b_{w,k}[0]=\pi[0]=0$, and $\sigma_z^2=\sigma_v^2=0.5$. 

\textit{GCI-based CUS estimation\ steps for each monitoring time instant }$t$, $t\geq 1$ and $\forall w,k$:  \textit{(i)}  retrieve (via the ME) the CUS measurements per workload and task type to establish $\tilde{b}_{w,k}[t-1]$; \textit{(ii)} perform the estimation of \eqref{eq:kalman_process_update}--\eqref{eq:kalman_process_update2};  \textit{(iii)}\ retain the value of the estimated CUS per workload via \eqref{eq:kalman_measurement_update} and \eqref{eq:r_w}.

\textbf{4. TTC confirmation and service rate per workload:} Let us assume that a reliable CUS estimation becomes available for workload $w$, $1\leq w\leq W[t_{\text{init}}]$,  at monitoring time instant\footnote{The practical method to determine $t_\text{init}$ is described in Section \ref{Sec5}.} $t_\text{init}$. The GCI can then confirm that $d_w[t_\text{init}]$ (the requested TTC\ for workload $w$ at $t_\text{init}$) is achievable by Dithen under appropriate adjustment of the workload \textit{service rate}, $s_w[t]$, for each monitoring time $t$, $t\geq t_\text{init}$. The service rate  $s_w[t]$ corresponds to the number of CUs allocated to workload $w$ for the time interval between monitoring instants $t$ and $t+1$. Fractional values (e.g.,   $s_w[t]=0.7$) indicate that one CU is allocated to workload $w$ for $s_w[t]\times100\%$ of the time between $t$ and $t+1$.  If the combination of  $d_w[t_\text{init}]$  with the workload CUS estimate leads to $s_w[t_\text{init}]>N_{w,\text{max}}$, with $N_{w,\text{max}}$ a predetermined CU upper limit ($\forall w$: $N_{w,\text{max}}=10$ in our experiments),   $d_w[t_\text{init}]$   is extended such that $s_w[t_\text{init}]=N_{w,\text{max}}$. This process confirms $d_w[t_\text{init}]$ (or its extension)\ as the TTC\ for workload $w$. 

The algorithm to determine   $s_w[t]$   for each workload $w$ and each $t\geq t_\text{init}$ is presented in Section \ref{Sec3} and is carried out by the GCI based on the estimated CUS per workload.
All LCIs of Dithen are given individual tasks from each workload $w$ according to   $s_w[t]$ by the GCI.

\section{Workload Execution with Confirmed TTC}\label{Sec3}

The GCI\ of Dithen ensures that each workload is executed within its remaining TTC by an allocation mechanism based on proportional fairness. The proportional fairness goal can then be stated as: at each monitoring instance \(t\) and for each workload $w\:\left( 1 \leq w \leq W[t] \right) $, the GCI maximizes an objective function of the service rate, $s_w[t]$, that ensures all workloads are served \textit{proportionally} to their CUS requirement, \(r_{w}[t]\) [given by \eqref{eq:r_w}], and \textit{inversely-proportionally} to their TTC, $d_w[t] $. The latter is defined via an appropriate SLA mechanism once a workload is submitted for execution and an initial workload CUS estimate becomes available. In this work, we adopt the objective function: 

\begin{equation}
%\begin{align*}
f(s_w[t]) = r_{w}[t]\ln(s_w[t]) -d_w[t]s_w[t].
%\end{align*}
\label{eq:f(p_w)}
\end{equation}

The subtraction in \eqref{eq:f(p_w)} contrasts between the workload's CUS requirement, $r_w[t]$, and the TTC requirement, $d_w[t]$. In addition, following proportional fairness problems of other resource allocation work (notably in cellular network scheduling algorithms \cite{Margolies2014fairness}), we opted for the use of the natural logarithm in the demand side of the objective function and pursue the maximization of  $f(s_w[t])$. Specifically, when the condition   $\sum_{w=1}^{W[t]} r_w[t]    \le c_{\text{tot}}[t]$ is satisfied, it is straightforward to show that the optimal solution to the maximization of \eqref{eq:f(p_w)}  is (\(\forall s_w[t] > 0\))

\begin{equation}
%\begin{align*}
s_w^*[t] = \text{arg} \; \text{max} \left\{ f \left( s_w[t] \right) \right\}=\frac{r_{w}[t]}{d_w[t]}.   
%\end{align*}
\label{eq:arg_max}
\end{equation}
This corresponds to the case where enough CUs are available to accommodate the demand and, therefore, allocation of service rates is carried out according to the required CUSs and TTC per workload at each monitoring time instant \(t\). We can then calculate the total required CUs for optimal operation as: 

\begin{equation}
%\begin{align*}
N_\text{tot}^{*}[t]=\sum _{w=1}^{W[t]}s_{w}^*[t]=\sum _{w=1}^{W[t]}\frac{r_{w}[t]}{d_w[t]}.   
%\end{align*}
\label{eq:N_tot_optimal}
\end{equation}

However, due to volatility in both workload submission and CU availability in Dithen, it is likely that, for most monitoring instances $t$,   $N_\text{tot}^{*}[t]$ differs from $N_\text{tot}[t]$ [the actual number of CUs, calculated by \eqref{eq:N_tot}]. In such cases, we can  adjust the optimal service rates of   \eqref{eq:arg_max} proportionally to the relative distance between   $N_\text{tot}^{*}[t]$ and $N_\text{tot}[t]$. Specifically, if $N_\text{tot}^{*}[t]>N_\text{tot}[t]+\alpha,$ with $\alpha$ the AIMD additive constant defined in the next section ($\alpha > 0$), we downscale the optimal service rate of each workload to:\

\begin{eqnarray}
\forall w: s^{-}_w[t] & = &\frac{r_{w}[t]}{d_w[t]} \left( 1-\frac{N_\text{tot}^{*}[t]-N_\text{tot}[t]-\alpha}{N_\text{tot}^{*}[t]}  \right) \nonumber \\ & = &\frac{ N_\text{tot}[t] + \alpha }{N_\text{tot}^{*}[t]} s_w^*[t] .   
\label{eq:s_w_scaledown}
\end{eqnarray}
If  $N_\text{tot}^{*}[t]<\beta N_\text{tot}[t]$, with $\beta$ the AIMD\ scaling constant defined in the next section ($0<\beta<1$), we upscale the optimal service rate of each workload  to:

\begin{eqnarray}
\forall w: s^{+}_w[t] & = & \frac{r_{w}[t]}{d_w[t]} \left( 1+\frac{\beta N_\text{tot}[t]-N_\text{tot}^{*}[t]}{N_\text{tot}^{*}[t]}  \right) \nonumber \\ & = & \frac{\beta  N_\text{tot}[t]}{N_\text{tot}^{*}[t]}s_w^*[t] .   
\label{eq:s_w_scaleup}
\end{eqnarray}
Finally, if $\beta N_\text{tot}[t] \leq N_\text{tot}^{*}[t] \leq N_\text{tot}[t]+\alpha$,  the service rates of    \eqref{eq:arg_max} are used. The use of $\alpha$ and  $\beta$ in \eqref{eq:s_w_scaledown} and \eqref{eq:s_w_scaleup} ensures the service rate adjustment is considering the possible additive increase or multiplicative decrease that may occur via the AIMD\ algorithm after the service rate allocation is established for the interval between $t$ and $t+1$.\ 

% \begin{figure}[t]
% %
% 
% 1 \ \ \ \ $l = 0$ \\
% 2 \ \ \ \ $s_w[l] =0$ \\
% 3 \ \  \ \ \texttt{end} $ = $ \texttt{FALSE}\\  
% 4 \ \ \ \ \textcolor[rgb]{0,0.501961,0.501961}{\% algorithm iterations }\\
% 5 \ \ \ \ \textbf{while } \texttt{end == FALSE} \textbraceleft   \\
% 6 \ \ \ \ \ \ \ \textbf{for } \texttt{[$w$ = 1; $w \leq W$; $w$++]}   \\
% 7 \ \ \ \ \ \ \ \ \ \ $s_w[l+1] = s_w[l]+h \times\ \left(\frac {r_{w}[t]}{s_w[l]}-d_w[t] \right) $\\
% 8.a\ \ \ \ \ \ \textbf{if } $\sum _{w=1}^{W[t]}s_w[l+1]  \geq N_\text{tot}[t]$\\ 8.b\ \ \ \ \ \ \ \ \ \ \ \ \ \ \ \ \ \ \ \ \textbf{or} $\text{max}_{\forall w}|s_w[l+1] - \frac {r_{w}[t]}{s_w[l]}|\leq s_\text{thres}$\textbraceleft \\
% 9 \ \ \ \ \ \ \ \ \ \ \texttt{end} $ = $ \texttt{TRUE} \\
% 10 \ \ \ \ \ \ \ \ \ $\forall w: s_{w}[t+1] = s_w[l]$ \textcolor[rgb]{0,0.501961,0.501961}{\% set practical service rate} \\
% 11 \ \ \ \ \ \ \ \ \ $\forall w: s_{w}^{*}[t+1] =\frac{r_{w}[t]}{d_w[t]}$ \textcolor[rgb]{0,0.501961,0.501961}{\% set optimal service rate} \\
% 12 \ \ \ \ \ \ \ \ \ $N_{\text{tot}}^{*} =\sum _{w=1}^{W[t]}s_w[t+1]$ \textcolor[rgb]{0,0.501961,0.501961}{\% set optimal CUs} \\
% 13 \ \ \ \ \ \textbraceright\ \\
% 14 \ \ \ \ \  $l \leftarrow l+1$ \\
% 15 \ \ \ \textbraceright\ \\
% 
% \caption{Proportional fairness algorithm for service rate allocation, with $h$ and $s_\text{thres}$ predetermined step size and threshold parameters, respectively. }
% \label{fig:PropFair}
%\end{figure}

  \section{Scaling with Additive Increase Multiplicative Decrease}\label{Sec4}

For any CaaS system, $N_\text{tot}^{*}[t]$ of \eqref{eq:N_tot_optimal} and $N_\text{tot}[t]$ of \eqref{eq:N_tot}   must be tightly coupled in order to ensure that the available compute-unit time can meet the service demand and TTC requirements at any instant. This is because, if $N_\text{tot}^{*}[t]$ is substantially higher than  $N_\text{tot}[t]$, the delay to complete pending workloads can increase significantly and workload TTCs may be violated. Conversely, when $N_\text{tot}^{*}[t]$ is  significantly smaller than  $N_\text{tot}[t]$,   several CUs may billed on the service unnecessarily. Therefore, and in conjunction with the fact that  billing comes in hourly increments in Amazon EC2 spot-instances, sudden surges or dips demand will have a detrimental effect in the delay or cost of the deployment of Dithen. Hence, the goal of GCI component of Dithen is to maintain the resource reservation and workload service rates at the correct level. To this end, we propose the AIMD algorithm of  Fig. \ref{fig:FigAIMD}. By controlling the additive and scaling constants, $\alpha$ and $\beta$ respectively, we can examine the behavior of Dithen under a wide variety of workload submissions. It should be noted that the corresponding problem of selecting which spot instances to terminate  in the event that $N_\text{tot}[t] > N_\text{tot}^*[t]$ is trivial: per instance type, the prudent action is always to terminate spot instances with the smallest remaining time before renewal.

\begin{figure}[t]
1 \ \textcolor[rgb]{0,0.501961,0.501961}{\% algorithm iterations for any monitoring time instant $t$} \

2 \ \textbf{if } $N_\text{tot}[t] \leq N_\text{tot}^{*}[t]$    \\
3 \ \ \ \ \texttt{incr} $ = $ \texttt{TRUE}\\ 
4  \ \textbf{else }    \\
5 \ \ \ \ \texttt{incr} $ = $ \texttt{FALSE}\\ 
6 \ \textcolor[rgb]{0,0.501961,0.501961}{\% application of AIMD to tune $N_\text{tot}$ for the next instant} \\
7 \ \textbf{if } \texttt{incr == TRUE} \\
8 \ \ \ \ $N_\text{tot}[t+1]= \min\{N_\text{tot}[t] + \alpha, N_\text{max}\}$ \textcolor[rgb]{0,0.501961,0.501961}{\% add more CUs}  \\
9 \ \textbf{else} \\
10 \ \ $N_\text{tot \\}[t+1] = \max\{\beta N_\text{tot}[t], N_\text{min}$\} \textcolor[rgb]{0,0.501961,0.501961}{\% remove CUs} \\
 
\caption{Proposed AIMD algorithm; $\alpha$ is a positive constant, $\beta$ is a constant such that $0 < \beta \le 1$,  and $N_\text{max}$ and $N_\text{min}$  are the upper and lower bounds for $N_\text{tot}[t]$.}
\label{fig:FigAIMD}
\end{figure}

We refer to the  work of Shorten \textit{et. al.} \cite{shorten2006positive} for details on the the stability and convergence properties of AIMD algorithms. A key aspect from their analysis is that fast convergence to an equilibrium state is achieved if $\beta$ is small and smoother transitions are expected if $\beta$ is close to unity  \cite{shorten2006positive}. After extensive experimentation, we opted for the values of   $\beta=0.9$ and $\alpha=5$,  which exhibit sufficiently-fast convergence while at the same time ensuring that CUs are not released prematurely. 

While the AIMD\ algorithm tunes the total CU value, $N_\text{tot}[t]$, it does not select which instance types to deploy out of the $I$ possible. As detailed in Appendix \ref{sec:Appendix}, the recent status of Amazon spot instance pricing provides for proportional increase of pricing according to the number of compute units per instance. Moreover, the single-CU instance type exhibits the minimum price volatility, thereby making it the safest instance type to use. Therefore, we opt to use only single-CU instances in our experiments, i.e., $I=1$ and $p_1=1$, which alleviates the problem of selecting amongst a variety of instance types. However, depending on the evolution of pricing data from the IaaS provider, future work will expand our results into a variety of instance types.

\section{Experiments}\label{Sec5}

In order to examine our proposals, we have deployed Dithen using single-CU \texttt{m3.medium} spot instances of Amazon EC2 (see Appendix  \ref{sec:Appendix} for more details). As discussed in Section \ref{Sec2}, each instance has a corresponding LCI that is given   new tasks to process once the GCI detects that it is idle. In addition, one reserved EC2 instance, serving as the GCI, calculates the Kalman filter estimates based on the CUS measurements per task.    Under predetermined TTC per workload (which is confirmed by Dithen after an initial CUS estimation becomes available for the workload), it then derives the service rate per workload  in fixed time periods (i.e., within 1--5 minute intervals), as described in Section \ref{Sec3}. This is communicated to the ME and the LCIs (see Fig. \ref{fig:FigDithen}). The GCI also carries out the AIMD algorithm of Section \ref{Sec4} in order to control the increase or decrease of spot instances according to the demand. The utilized AIMD parameters for all experiments were set to: $\alpha=5$,  $\beta=0.9$, $N_\text{min}=10$, $N_\text{max}=100$ and $\forall w$: $N_{w,\text{max}}=10$ (maximum service rate per workload). A SQL database is used by the ME to keep track of the tasks completed per workload. Finally, the produced results, as well as a summary of the intermediate progress, is communicated to the
user by the web interface of the FE (Fig. \ref{fig:FigFrontEnd}). 
\subsection{Utilized Workloads}

\begin{figure}[t]
\begin{center}
\includegraphics[scale=0.10]{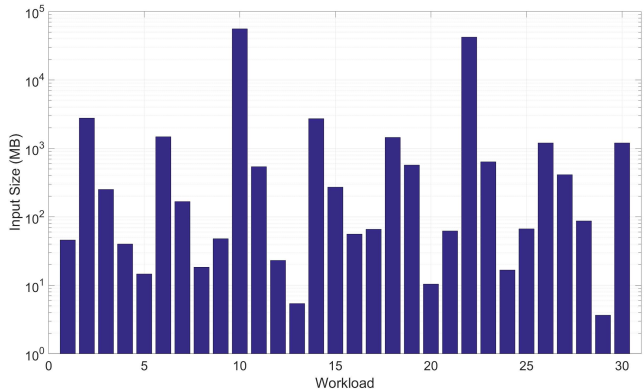}
\end{center}
\caption{Size of inputs for each of the thirty workloads used in our experiments.}
\label{fig:FigWorkloads}
\end{figure}

All multimedia inputs, processing scripts and executable files are placed on Amazon S3 via the  uploading service (``add'' button) available within the FE of Dithen [Fig. \ref{fig:FigFrontEnd}(a)]. Thirty different workloads, each with a  random number of tasks were used in our experiments. Eight of the workloads were scripts running the Viola-Jones  classifier \cite{viola2004robust} for face detection in  images. The range of possible values for the number of inputs (i.e., images or videos) for these workloads
 was between 1 and 1000. Eight of the workloads were scripts using FFMPEG to transcode videos to different bitrates via a variety of codecs \cite{garcia2010study}. Each workload had between 1 and 20 videos to transcode, and we also added two large transcoding workloads with 200 and 300 videos. These were used to examine
 the responsiveness of the Dithen system under sudden spikes of demand. Seven of the workloads were using the OpenCV BRISK keypoint detector and descriptor extractor \cite{ leutenegger2011brisk}. Finally, seven workloads used the Scale Invariant Feature Transform (SIFT) salient point descriptor \cite{lowe2004sift}, which was deployed as compiled Matlab code with the Mathworks \texttt{deploytool}. The total size of the inputs per workload is given in Fig. \ref{fig:FigWorkloads}.
Workloads were introduced once every five
 minutes in the order depicted in Figure \ref{fig:FigWorkloads}.

\subsection{Performance of Kalman-based CUS Estimation }

The proposed Kalman-based CUS\ estimation process of Section \ref{Sec3} is compared against the  ``ad-hoc'' estimator that carries out the CUS\ estimation of \eqref{eq:kalman_measurement_update}, albeit with the scaling coefficient being set to the fixed value: $\kappa_{w,k}[t]=0.1$, which was shown to perform best amongst other settings. Moreover, as an external comparison, we also utilize the well-known second-order autoregressive moving average (ARMA) estimator of Roy \textit{et. al.}  \cite{roy2011efficient}  that has been shown to perform well for workload  forecasting. ARMA estimates the CUS\ required to complete a workload at time $t+1$ via  

\begin{eqnarray}
\hat{b}_{w,k} [t+1] & = & \delta \times b_{\text{norm},w,k} [t]+ \gamma \times  b_{\text{norm},w,k} [t-1] \nonumber\\
 & + & ( 1-\delta-\gamma )  \times  b_{\text{norm},w,k} [t-2],   
\label{eq:ARMA_CUS_pred}
\end{eqnarray}
where: $b_{\text{norm},w,k} [t,t-1,t-2]$ are calculated by summing the total execution time of media type $k$ of workload $w$ at times $t,t-1,t-2$ and dividing it by the percentage of the workload that has been completed until then; and $\delta$ and $\gamma$ are scalars having the values recommended by  Roy \textit{et. al.}  \cite{roy2011efficient}. We chose ARMA as the most suitable benchmark because other workload forecasting methods (like  the ARIMA model   \cite{debusschere2012hourly,Calheiros2014arima}) require extensive past measurements from previous executions of other workloads, as well as a long sequence of measurements in order to produce reliable estimates, thereby making them unsuitable for our case.  

\begin{figure}[t]
\begin{center}
\includegraphics[scale=0.31]{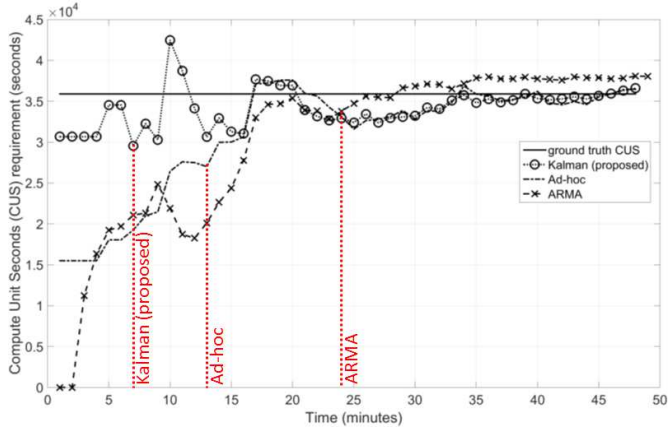}
\end{center}
\caption{Example of the convergence of various CUS estimation methods for the case of an FFMPEG\ workload under 1-min monitoring interval.   }
\label{fig:EstimatorsSlope}
\end{figure}

\begin{figure}[t]
\begin{center}
\includegraphics[scale=0.31]{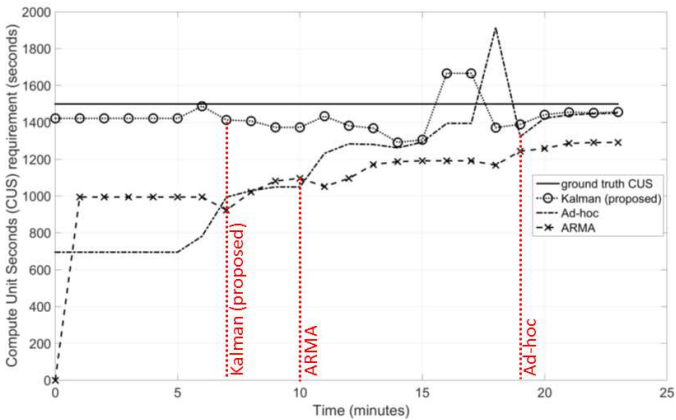}
\end{center}
\caption{Example of the convergence of various CUS estimation methods for the case of a SIFT\ descriptor (Matlab-based) workload under 1-min monitoring interval.}
\label{fig:EstimatorsSlope2}
\end{figure}

Two representative examples of the convergence behaviors of all methods under comparison are given in Fig. \ref{fig:EstimatorsSlope} and Fig. \ref{fig:EstimatorsSlope2}.  As illustrated in the figures, the Kalman and ad-hoc estimator exhibit an underdamped behavior until convergence.  We can therefore use the slope of the CUS\ estimation across time to determine the monitoring time instant $t_\text{init}$ when the proposed  Kalman and the ad-hoc estimator can provide a reliable CUS\ estimation per workload and task type. Specifically, when the slope of the CUS\ estimation becomes negative for the first time, each estimator establishes a CUS\ estimate for each workload with acceptable accuracy.  However, ARMA\ does not exhibit such underdamped behavior, since it is a moving-average based estimator. Therefore, we relied on a conventional convergence detection criterion for ARMA: when the ARMA\ estimate deviation within the window of the last three measurements is found not to exceed 20\% from the mean value derived from the values of the window (ten measurements are used for the case of 1-min monitoring), we determine that the estimate is reliable enough to be used. The setup for the window size and variability threshold was selected after testing with a variety of possible values. In the examples of  Fig. \ref{fig:EstimatorsSlope} and Fig. \ref{fig:EstimatorsSlope2}, the time instant when each method reaches its reliable estimate under the described setup is marked with the red dotted vertical lines.   

Table \ref{tab:TabTTCest} presents the average time each estimator took to reach its CUS estimate for each workload type, as well as the CUS\ percentile mean absolute error (MAE). The summary over all workloads (per monitoring interval) is given at the bottom of the table. Evidently, the proposed Kalman-based approach reduces the average time to reach a reliable estimate by more than 20\% in comparison to the other estimators and is found to be the quickest estimator in all but one case. At the same time, the proposed estimator attains comparable  accuracy to the ad-hoc estimator and is found to be significantly superior to ARMA. This is especially pronounced in the case of the 1-minute monitoring, where the use of the proposed Kalman-based approach instead of an ARMA approach provides for 38\% reduction in estimation time and decreases the average estimation error from 16.4\% to 4.5\%. This indicates that, under the usage of the proposed CUS\ estimator and 1-minute monitoring, the GCI\ is expected to have reliable estimates per workload (and thereby confirm that its requested TTC is achievable) within 6--11 minutes from its launch. 
Finally, when we compare the performance of one-minute monitoring to five-minute monitoring, Table \ref{tab:TabTTCest} shows that increase in the measurement granularity results in significant improvement in both the accuracy and time required to reach a reliable estimate. Specifically, for the proposed Kalman estimator, increased monitoring frequency reduces the the average estimation time by 44\% and reduces the overall MAE from 13.1\% to 4.5\%.

\ 

\begin{table*}
\caption{Average time to reach CUS  estimation per type of workload and percentile Mean Absolute Error (MAE)\ of the derived estimate. The last column presents the percentile time reduction when switching from 5-min monitoring to 1-min monitoring intervals. The best result per category is indicated in boldface font.}

\begin{center}
\begin{tabular}{|c|c|c|c|c|c|}
\hline 
Control interval & \multicolumn{2}{c|}{5-min monitoring } & \multicolumn{2}{c|}{1-min monitoring }& Time 

Reduction (\%) by going \tabularnewline
 
\textbf{ Face Detection} &  Time & MAE (\%)\ & Time & MAE (\%) &  from 5-min to 1-min monitoring\tabularnewline
\hline 
\hline 
Kalman-based & \textbf{13m 45s} & 5.6 & \textbf{10m 38s} & \textbf{4.6} & 22.7\textbf{}   \tabularnewline
\hline 
Ad-hoc  & 28m 08s & \textbf{4.5} & 17m 53s  & 5.3 & 36.4  \tabularnewline
\hline
 ARMA & 23m 08s & 22.1 & 12m 08s & 27.8 & 47.6\tabularnewline
\hline
\hline 
\textbf{Transcoding} & Time & MAE (\%) &Time & MAE (\%) & \tabularnewline
\hline 
\hline 
Kalman-based & \textbf{16m 53s} & 14 & \textbf{07m 54s}  & 7.8 & 53.2    \tabularnewline
\hline 
Ad-hoc  & 26m 53s & \textbf{8.9} &  10m 36s  & \textbf{1.5} & 60.6   \tabularnewline
\hline 
ARMA & 28m 08s & 13.9 & 18m 45s & 18.1 & 33.4  \tabularnewline
\hline
\hline 
\textbf{Feat. Extraction} & Time & MAE (\%) &Time & MAE (\%) &  \tabularnewline
\hline 
\hline 
Kalman-based & \textbf{13m 34s} & 12.1 & 11m  54s&  \textbf{1.4} & 12.3 \ \tabularnewline
\hline 
Ad-hoc  & 18m 34s & 6.4 & 20m 24s  &1.9 &-9.9  \tabularnewline
\hline
ARMA & 20m 43s & \textbf{5.7} & \textbf{11m 09s} & 12.1 & 46.2 \tabularnewline
\hline
\hline 
\textbf{SIFT} & Time & MAE (\%) & Time &MAE (\%) & \tabularnewline
\hline 
\hline 
Kalman-based & \textbf{21m 26s} & 20.6  & \textbf{06m 18s} & 4.1 & 70.6  \tabularnewline
\hline 
Ad-hoc  & 23m 54s & 18.9 & 08m 06s & \textbf{0.1} & 66.1 \tabularnewline
\hline
ARMA & 20m 00s & \textbf{20.1}  & 15m 00s & 7.6 & 25.0\tabularnewline
\hline
\hline 
\textbf{Overall Average} & Time & MAE (\%) & Time & MAE (\%) & \tabularnewline
\hline 
\hline 
Kalman-based & \textbf{16m 25s} & 13.1  & \textbf{09m 11s} & 4.5 & 44.1  \tabularnewline
\hline 
Ad-hoc  & 24m  22s & \textbf{9.7} & 14m 15s & \textbf{2.2} & 34.6 \tabularnewline
\hline 
ARMA & 23m 00s & 15.5 & 14m 15s & 16.4 & 38.0\tabularnewline
\hline

\end{tabular}
\end{center}
\label{tab:TabTTCest}
\end{table*}

\subsection{Results for Cumulative Cost of Workload Execution }  
\label{sec:CumulativeCost}

We now investigate the management of spot instances so that each workload is completed under a fixed TTC that is sufficiently large to allow for fluctuation in the number of utilized instances.  

As external comparisons, our first choice is Amazon's Autoscale service (termed as ``Amazon AS''), which is widely deployed in practice  \cite{tighe2014autoscale}. Amazon AS does not carry out CUS estimation or TTC-abiding execution, and one can  only control the number of instances based on CPU utilization and bandwidth constraints. Therefore, under these conditions, we configured all workloads to execute within an Amazon AS group that examines the average CPU usage at all utilized CUs in five-minute intervals. If the group detected that the average CPU utilization was more than 20\%, new instances were started\footnote{After extensive experimentation, the value of 20\% was found to provide for the best results with Amazon AS. This is because average\ utilization values between 18\%\ and 22\% represent the average CPU\ usage observed within active time intervals when an instance alternates between downloading files (2\%--10\%\ CPU\ utilization) and actually  executing a compute-intensive task (close to 100\% CPU\ utilization).}. Otherwise, Amazon AS terminated some of the active instances. We then executed all workloads in Amazon AS and measured the longest time to complete a workload under two scaling policies. The first represented a conservative approach where reducing the execution time is not of critical importance. In this case, a single instance is added or removed when a monitoring interval occurs. The longest completion time was found to be 2 hr 7min. The second scaling policy started and stopped ten instances instead of one, to represent a scenario where reduced execution time is of importance. In this case, the longest time to complete a workload was found to be 1 hr and 37 min. Both of these times were then used as the two fixed TTC settings for all workloads in Dithen. 

Beyond Amazon AS, in order to benchmark our AIMD-based scaling of Fig. \ref{fig:FigAIMD} against other alternatives for CU adjustment, we utilized  the mean-weighted-average and linear-regression methods   of Gandhi, Krioukov \textit{et. al.}  \cite{Gandhi2012autoscale,krioukov2011napsac} (termed ``MWA'' and ``LR'', respectively) to set the number of CUs for the next monitoring interval, $N_\text{tot}[t+1]$. We selected MWA and LR for our comparisons because previous work  \cite{Gandhi2012autoscale} has shown them to be amongst the most accurate predictive resource controllers. Both MWA and LR utilized the proposed Kalman-based CUS estimation process and the service rate allocation of  \eqref{eq:N_tot_optimal} to determine when to increase or decrease CUs. Specifically: \textit{(i)} MWA\ sets the number of CUs via

\begin{equation}
%\begin{align*}
N_\text{tot}[t+1]=\frac{1}{6} \sum_{i=t-5}^{t}N_\text{tot}^{*}[i],   
%\end{align*}
\label{eq:MWA_CUs}
\end{equation}
where $N^*_\text{tot}$ is the optimal number of CUs derived via \eqref{eq:N_tot_optimal} for each monitoring time instant; \textit{(ii)} LR\ sets $N_\text{tot}[t+1]$ to be the result of extrapolating the line derived via linear regression from $\{N_\text{tot}^*[t], \dots, N_\text{tot}^*[t-5]  \}$ (current plus five previous CU settings). Finally, in order to see the performance of the direct-compensation approach, we also utilized the case where no filtering or other adjustment is being used and we simply set  $N_\text{tot}[t+1]=N_{\text{tot}}^{*}$   (termed as ``Reactive'').

% \begin{figure}[t]
% \begin{center}
% \includegraphics[scale=0.7]{TTCs.pdf}
% \end{center}
% \caption{TTCs of each workload for the proportional experiment.}
% \label{fig:FigTTCs}
% \end{figure}

\begin{figure}[t]
\begin{center}
\includegraphics[scale=0.31]{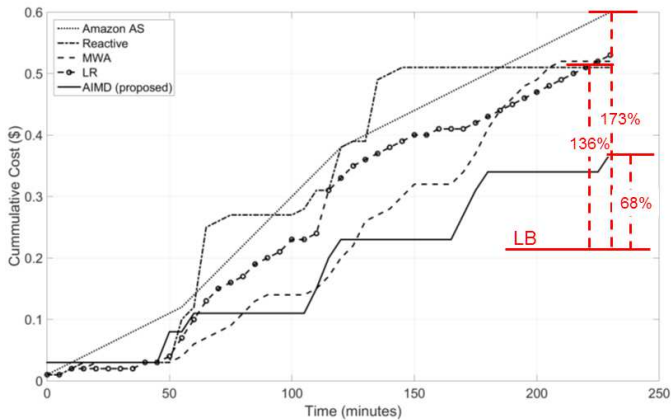}
\end{center}
\caption{Cumulative cost of processing all workloads of Fig. \ref{fig:FigWorkloads} under fixed TTC of  2 hr 7 min per workload. LB indicates the lower bound.}
\label{fig:Cost}
\end{figure}

\begin{table*}[t]
\caption{Overall cost of different methods and comparison against the proposed method and the lower bound (LB). }

\begin{center}
\begin{tabular}{|c|c|c|c|c|c|c|}
\hline
System & AIMD (proposed)  & Reactive & MWA & LR & AS & LB \tabularnewline  
\hline
Overall cost (\$) & 0.41 & 0.51 & 0.52 & 0.53 & 1.02 & 0.22 \tabularnewline \hline
Average cost reduction of proposed vs. other methods (\%) & -- & 20 & 21 & 23 & 60 & -- \tabularnewline
\hline
Average cost increase vs. LB (\%)\ & 86 & 132 & 136 & 141 & 364 & -- \tabularnewline
\hline 
Max. \#\ of instances at any time by each method & 13 & 28 & 21 & 24 & 91 &-- \tabularnewline \hline
\end{tabular}
\end{center}
\label{tab:TabCostSummary}
\end{table*}

\begin{figure}[t]
\begin{center}
\includegraphics[scale=0.31]{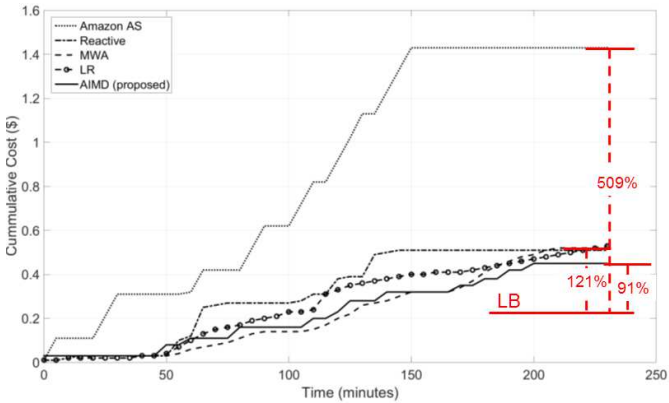}
\end{center}
\caption{Cumulative cost of processing all workloads of Fig. \ref{fig:FigWorkloads} under fixed TTC of  1 hr 37 min per workload. LB indicates the lower bound.}
\label{fig:CostLowDelay}
\end{figure}

Figure \ref{fig:Cost} and Figure \ref{fig:CostLowDelay} show the cumulative cost of each approach during the course of both experiments with the two TTC values. Evidently, the cost of Amazon AS is  significantly higher than that of all other approaches. This is primarily because the Amazon AS is the only approach that does not use CUS estimations and instead bases its decisions solely on CPU\ utilization. Therefore, it continues to scale up the number of instances even when it is nearing completion of the workloads' processing and only scales down after workloads have been completed and CPU utilization decreases due to inactivity.   

Amongst MWA, LR and Reactive, MWA is superior as it incurs less cost for the majority of the experiment (and, as expected, Reactive is the worst). However, all three methods end up incurring very comparable cost for the completion of all workloads. Interestingly, Reactive turns out to be (marginally)\ the cheapest of the three for this experiment even though it uses the largest number of instances of the three methods at one point. The reason for this is that, while Reactive scales up very quickly it also scales down rapidly and, for this particular experiment, this behaviour worked in its favour. However, this is not expected to be always the case, as Reactive does leave many instances  idle for a large portion of their billed time.

 The proposed AIMD-based scaling initially scales up when it detects the large workloads, then maintains this level, and then begins to scale down as it nears the experiment completion. For the experiments of Figure \ref{fig:Cost}, this leads to overall savings of 30\% against MWA, 29\% against LR, 27\% against\ Reactive and 38\%\ against Amazon AS. For the experiments of Figure \ref{fig:CostLowDelay}, the equivalent savings were:\  14\%, 15\%, 12\%\ and 69\%\footnote{{It should be noted that the controller does incur some overhead cost. If we were to subtract the cost from Amazon AS (Reactive, MWA and LR also require a controller and thus have the same overhead as AIMD) it would not improve its  performance by no more than 5\%, with this percentage diminishing as the workload size increases. Finally, it is also important to note that the controller instance does not have to run in AWS; instead, it could operate under a captive computing environment, thereby incurring no billing cost from the cloud provider.}}. Overall, beyond the advantage of providing for scaled-up execution under TTC constraints, the 38\%--69\% savings demonstrated in Figure \ref{fig:Cost} and Figure \ref{fig:CostLowDelay}  allow for significant profit margin for cloud\ service providers that would deploy large-scale multimedia applications via the techniques used in Dithen, versus utilizing Amazon AS\ directly. 

The overall savings for both experiments, as well as the maximum number of instances used by the proposed algorithm against all other benchmarks are summarized in  Table \ref{tab:TabCostSummary}. It should be emphasized that, beyond the cost savings, all the workloads in the proposed AIMD approach finished before their execution time exceeded the predetermined TTC of each experiment. Such TTC-abiding execution is a significant feature that Amazon AS cannot provide.

Finally, the bottom right of Figure \ref{fig:Cost} and Figure \ref{fig:CostLowDelay} includes a red horizontal line indicating the estimated billing if all workloads would be processed such that all billed instances would be occupied 100\%\ of the time. This constitutes the lower bound for the billing cost (termed ``LB'') as no operational approach can achieve lower cost. Evidently, the proposed approach incurs 68\%--91\%\ higher cost than LB, but all other approaches incur 135\%--510\%\ higher cost than LB. It should be noted that both the LB and all the examined approaches include the delay to transport of data to and from the instances. If this would be removed, all costs would be lowered by approximately 27\%. Overall, the results of Figure \ref{fig:Cost} and Figure \ref{fig:CostLowDelay}   demonstrate that the proposed AIMD-based scaling of CUs is a simple and effective method towards approaching the lowest possible cost incurred from the cloud computing infrastructure, while at the same time satisfying the TTC\ constraint of each workload. \ \

% \begin{figure}[t]
% \begin{center}
% \includegraphics[scale=0.7]{MaxInstances.pdf}
% \end{center}
% \caption{Maximum number of CUs for all methods for the experiments with fixed and proportional TTC. \textbf{\textless\textless NEED\ TO\ ADJUST\ THE\ FIG A\ BIT\textgreater\textgreater}\ }
% \label{fig:MaxInstances}
% \end{figure}

\subsection{Comparison Against Amazon Lambda}

Recently, Amazon begun offering its own CaaS service for the execution of Javascript code via its Lambda service. Despite this being more limiting due to the inefficiency of Javascript code, we compared the cost of running three large Javascript-based workloads on Dithen and Lambda. In this experiment we ran ``blur'', ``rotate'' and ``convolve'' operations from the Javascript version of the widely-used  ImageMagick image manipulation program \cite{imagemagick}. We chose these functions as they represent a cross section of computational requirements of the various ImageMagick functions. Each function was executed on 25,000 images encompassing a wide variety of sizes and pixel counts. We also opted for the 1024MB-memory configuration for all Lambda functions to avoid any memory bottlenecks during execution. Again, Dithen was tuned to match the execution time of each workload in Lambda. This was done because the latter is dependent on how quickly requests can be sent to call the functions through the Amazon Web Service Command Line interface (or any other such API), while the execution time for workloads in Dithen is completely tunable based on their specified TTC. This flexibility of TTC-abiding execution\ per workload is an advantage of our proposal against Lambda.

A comparison of the cost of executing the workloads is given in Table \ref{tab:LambdaComparison}. It is interesting to notice that, as the run time of the function decreases, Lambda becomes a more viable option. For example,  the average cost of running the most compute-intensive function (Blur function in Table \ref{tab:LambdaComparison})  was 3.34 times higher on Lambda than it was on Dithen. In contrast the average cost of running the fastest and least compute-intensive function (the rotate function) was found to be slightly less on Lambda than on Dithen. {This result can be understood as follows. AWS Lambda allocates cores based on the memory consumption. For example, if the Lambda functions run on an EC2 instance with 4 GB memory and 2 cores and the functions require only 1 GB of memory, Lambda will allocate only $\frac{1}{4} \times 2$ cores, so it won't utilize the full processing power of the instance, thereby making the functions run longer. This implies that, when Lambda handles low-load tasks (i.e., easily executable even when a non-dedicated core is available),  it becomes advantageous to Dithen. However, when high-load tasks are executed, the pricing and core allocation of Lambda becomes less advantageous since the execution time of complex tasks is significantly prolonged in comparison to Dithen (which always allocates an entire core per task, regardless of the task's complexity). Therefore, beyond simple web front-end type of tasks (which is the ideal application domain for AWS Lambda---hence its design and pricing being built around this), Lambda is not advantageous for the vast majority of more advanced computing tasks handled by a more generic and extendable CaaS platform like Dithen.} Overall, we were able to run the workloads on Dithen at more than 2.5 times lower cost (60\% reduction)\ in comparison to Amazon Lambda. This provides for substantial profit margin for a cloud service provider to deploy a large-scale multimedia application via the proposed approach instead of Lambda.

\begin{table}
\caption{Average cost of ImageMagick functions per image of the  25,000 dataset for Dithen and Amazon's lambda. }

\begin{center}
\begin{tabular}{|c|c|c|c|}
\hline
Function & Lambda Cost (\$) & Dithen Cost (\$) & Ratio \tabularnewline
\hline
Blur & $4.74 \times 10^{-5}$ & $1.42 \times 10^{-5}$ & 3.34\ \tabularnewline
\hline
Convolve & $1.68 \times 10^{-5}$ & $6.05 \times 10^{-6}$ & 2.78 \tabularnewline
\hline
Rotate & $5.5 \times 10^{-6}$ & $6.8 \times 10^{-6}$ & 0.81 \tabularnewline
\hline
Overall Average & $2.32 \times 10^{-5}$ & $9.20  \times 10^{-6}$ & 2.52 \tabularnewline
\hline

 \end{tabular}
\end{center}
\label{tab:LambdaComparison}
\end{table}

\subsection{Deep Learning and Split-Merge Workloads}
\begin{figure}[t]
\begin{center}
\includegraphics[scale=0.3]{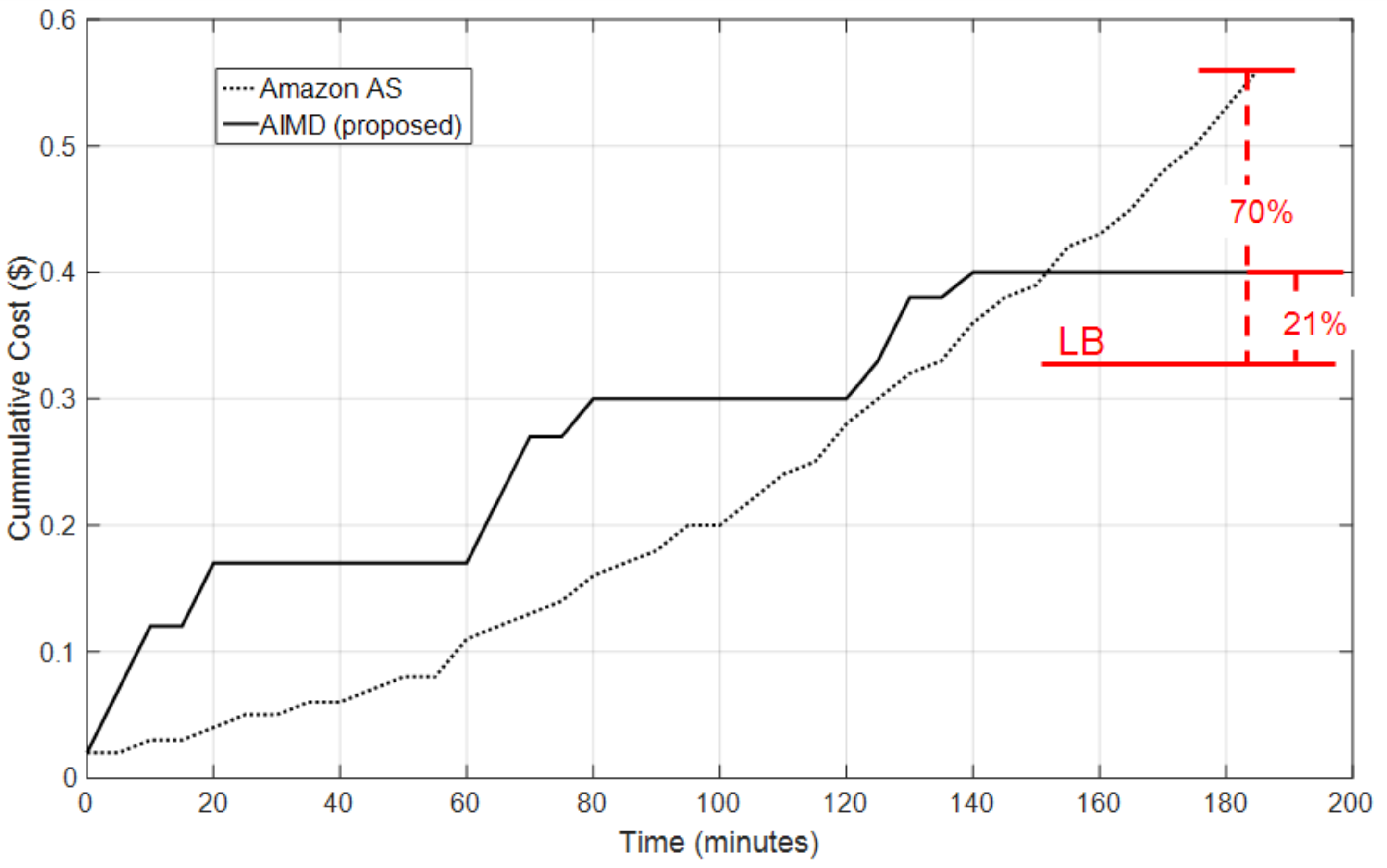}
\end{center}
\caption{Cumulative cost of an image classification workload based on deep CNNs. LB indicates the lower bound.   }
\label{fig:ImageCost}
\end{figure}

\begin{figure}[t]
\begin{center}
\includegraphics[scale=0.3]{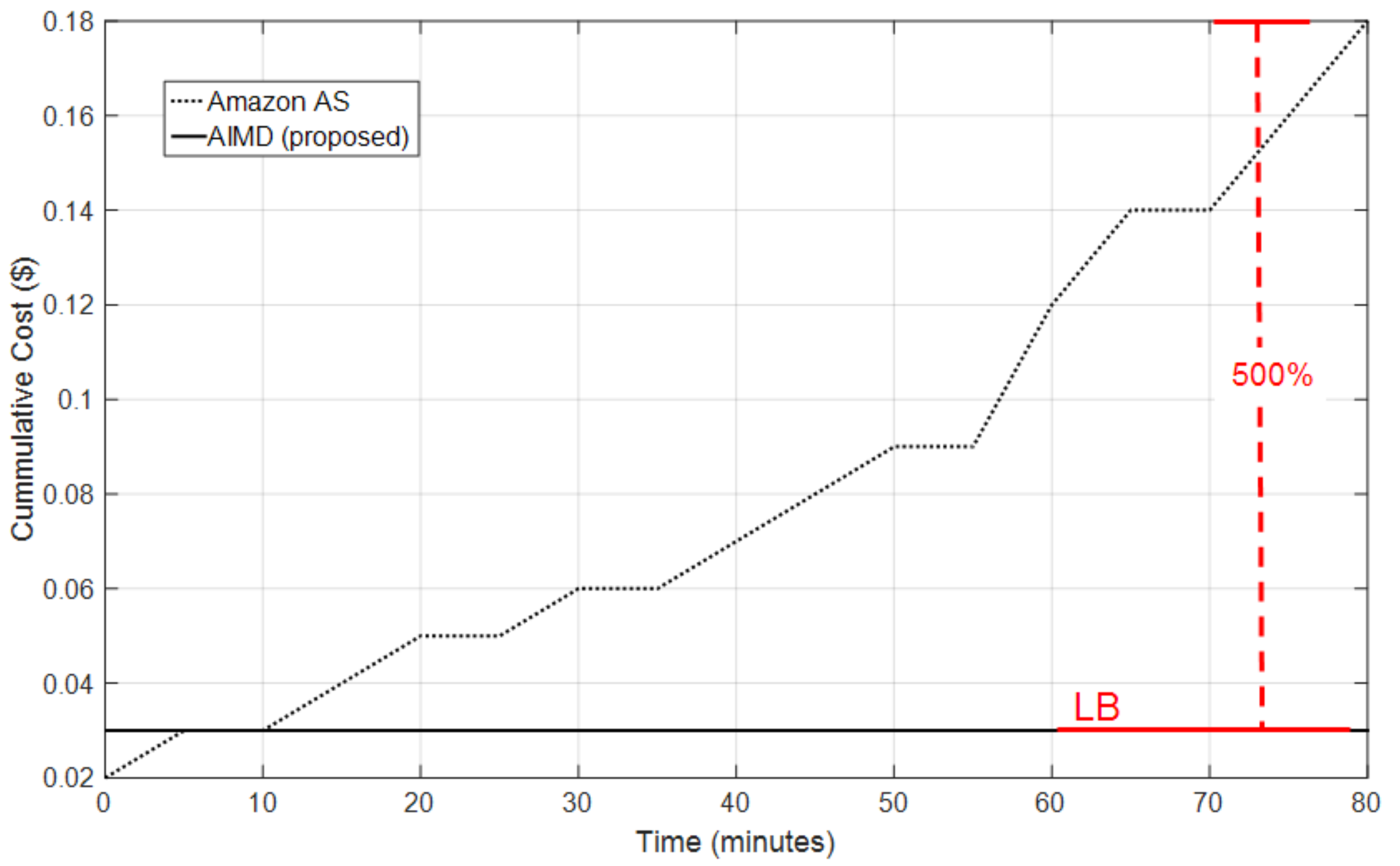}
\end{center}
\caption{Cumulative cost of a word histogram calculation workload based on Split-Merge (the workload is the standard one used within MapReduce testing \cite{Dean2008MapReduce}). LB indicates the lower bound.}
\label{fig:HadoopCost}
\end{figure}

We conclude our experiments by examining the performance of our platform when processing more complicated workloads, namely: \textit{(i)} an image classification application based  on a group of deep convolutional neural networks (CNN) \cite{chatfield2014return} that have been trained on ImageNet \cite{krizhevsky2012imagenet}\ and \textit{(ii)} a large-scale word histogram calculation, which is the standard example used with MapReduce-type of processing \cite{Dean2008MapReduce}. 

The first example is a representative case for a Split-Merge workload since multiple deep CNNs are used to classify each input image during the split stage and their results are aggregated via a voting process in the merge stage in order to produce the final classification result per image \cite{krizhevsky2012imagenet}. The inputs used for this workload comprised all images of the ``Holidays'' dataset \cite{chadha2015region}, as well as 50,000 additional images from ImageNet.

In the second example, the  workload measured the number of occurrences of words in a text file and then this was aggregated into a word histogram by a separate Reduce (or merge) instance. This is similar to a number of text based workloads where text data is analysed in order to gain insights into trends in market sentiment. The inputs used to test this workload were a selection of the Project Gutenberg \cite{Hart1997Gutenberg} library which was approximately 14,000 text files and 5.5GB of data. This example is used in order to demonstrate that, while Dithen is more amenable to multimedia workloads (where  the partitioning is inherent), it can also be used for more general workloads such as market sentiment analysis and the semantic analysis of text. {It should be noted that, beyond testing, Dithen could be used for deep learning \textit{training} workloads as well. For example Tensorflow could be used with batches of training sets and   the results of such batch-based training could be merged at a later stage, after several iterations have been carried out in batch mode. We plan to report on such experiments in a future paper.}

The experiments were invoked via the front end following the process described in Section \ref{sec:FE} and experimental benchmarking of the incurred cost occurred as described in Section \ref{sec:CumulativeCost}. Specifically, the workloads were first executed using Amazon's Autoscaling service (commonly used for such systems \cite{kouki2013scaling}), which was used to determine the TTC to use for our platform\footnote{For simplicity and brevity in our exposition, we did not include the results with the remaining methods in our presented comparisons (MWA, LR and Reactive), as they were found to incur similar overhead as in the previous experiments. Furthermore, no results are presented for Amazon Lambda, as Lambda cannot support such complex processing tasks.  } (since Amazon AS does not allow for TTC-abiding execution). Based on this process, the TTC was set to 1 hr 35 min for the first example and 1 hr 05 min for the second example. {In order to account for the time for the Merge step of each of the two workloads, the TTC\ for each Split stage was set to 90\% of the overall TTC.}\  

The cumulative cost of the image classification workload can be seen in Figure \ref{fig:ImageCost}. Similar to previous examples, the cost of Amazon AS is 38\%\ higher than the cost of the AIMD approach of Dithen. We can also see that the cost of this workload in Dithen is only 21\% higher than the lower bound, while the cost of the Amazon AS approach is 70\% higher than the lower bound. 

{The cumulative cost of the word histogram calculation workload is depicted in Figure \ref{fig:HadoopCost}. In this case, the cost of Amazon AS\ is six times that of the Dithen platform and  the lower bound. Interestingly, in this case, the cost incurred by the AIMD approach of Dithen is extremely close to the lower bound (less than \$0.005 higher) and remains constant at 3 cents.  This result is achieved because, in this particular case, Dithen was able to quickly and reliably identify the CUSs required to complete the Split tasks and determined that 3 spot instances  suffice for the completion of the workload within the predetermined TTC, and below the 1 hour mark (at which point additional charges are levied by AWS). Therefore, it avoided the unnecessary launch of new instances and its cost remained constant at 3 cents since the Split-Merge workload execution finished in 55 minutes. 

Overall, these examples show that the platform can be used to substantially lower the execution costs of complex workloads and, in certain circumstances, it is even possible for Dithen to approach the lower bound for the incurred cost.   

\section{Conclusions}\label{Sec6}

We present \textit{Dithen}, a novel Computation-as-a-Service framework, which supports the direct upload and execution of multimedia processing workloads. The Dithen architecture comprises multiple spot instances that execute tasks within the workloads until their compute units are fully utilized. Dithen uses the Additive Increase Multiplicative Decrease (AIMD)\ algorithm for the allocation or termination of compute units and a Kalman-based estimator for the required compute-unit-seconds for each type of task with each workload. Experiments based on Amazon EC2 spot instances demonstrated that, unlike all existing Platform-as-a-Service and Software-as-a-Service frameworks, Dithen provides for extreme scaling of commodity multimedia computing tasks (like large-scale transcoding, face detection and feature extraction workloads), without requiring any modification in the users' code base, and at substantially-reduced cost against all other alternatives. Moreover, unlike other services, Dithen allows for execution under time-to-completion constraints. The baseline form of the proposed service is available at {http://www.dithen.com} under the ``AutoScale'' option.\

\appendices 

\section{\label{sec:Appendix}}

We briefly analyze the computation costs of Linux instances on AWS EC2, as EC2 is considered to be the largest public cloud service provider today  \cite{Choy2014AWSLargest}  and our system is tested and deployed on the EC2 infrastructure. A comparison of the cost and EC2 compute units (ECUs) of various instance types is given in\footnote{Table \ref{tab:CostComparison} does not include all instance types available on Amazon's EC2. However, all non-included instances are memory, computation or storage variants of the instances depicted in Table \ref{tab:CostComparison}.} Table \ref{tab:CostComparison}. An ECU is defined as ``equivalent CPU capacity of a 1.0-1.2 GHz 2007 Opteron or 2007 Xeon processor''. The  \texttt{m3.medium}  instance (utilized in this paper)\ is a single CU instance with clock speed of 3.0--3.6GHz. From the table we can also see that the larger instances consist of increasing numbers of CUs (i.e., virtual cores available for computations) with similar clock speeds. We can also see that the ``On Demand'' cost and spot prices are both linearly-dependent on  the number of CUs. Thus, we can conclude that it is more efficient to use a large number of cheaper instances than small number of more expensive instances, as it allows for greater granularity when controlling the number of active instances without any corresponding increase in cost.

\begin{table*}
\caption{Cost of Various Linux Instances on the Amazon EC2 Platform in the North Virginia Region }

\begin{center}
\begin{tabular}{|p{4.5cm}|c|c|c|c|c|c|}
\hline
\textbf{Instance Type} & m3.medium & m3.large & m3.xlarge & m3.2xlarge & m4.4xlarge & m4.10xlarge\tabularnewline
\hline
\hline
EC2 compute units (ECUs) & 3  & 6.5 & 13 & 26 & 53.5 & 124.5\tabularnewline
\hline
CUs  & 1 & 2 & 4 & 8 & 16 & 40\tabularnewline
\hline
On-demand cost (\$) & 0.067  & 0.133 & 0.266 & 0.532 & 1.008 & 2.52\tabularnewline
\hline
Spot  price (\$)\ & 0.0081 & 0.0173 & 0.0333 & 0.066 & 0.1097 & 0.5655 \tabularnewline
\hline
Cost reduction when using spot  (\%)& 88 & 87 & 87 & 88 & 89 & 78  \tabularnewline
\hline

 \end{tabular}
\end{center}
\label{tab:CostComparison}
\end{table*}
\begin{figure}[t]

\includegraphics[scale=0.7]{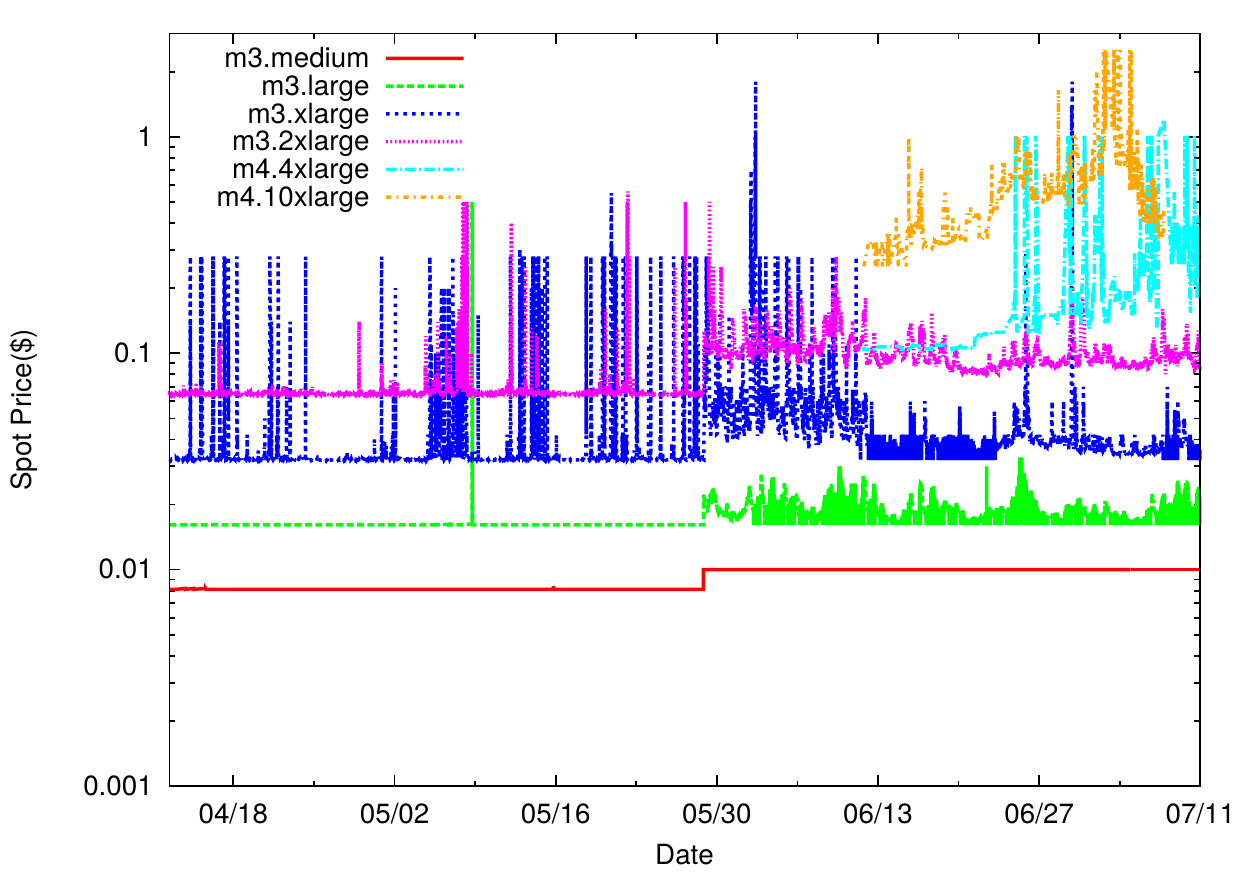}
\centering

\caption{Spot Price for various instance types from 11\textsuperscript{th} of April to the 11\textsuperscript{th} July 2015 }
\label{fig:FigSpot}
\end{figure}

From Table \ref{tab:CostComparison} we can also see the difference between the ``On Demand'' cost and the Spot price\footnote{The Spot prices depicted in Table \ref{tab:CostComparison} were taken on the 10\textsuperscript{th} July 2015.}. Spot instances are instances that will only function when a user's bid is greater than the current spot price. Essentially, the user gives up certainty of having computational resources available, in exchange for a significant reduction in the cost. We can see from Table \ref{tab:CostComparison} that this reduction ranges from 78\% to 89\%. However, it is difficult to run a CaaS service without guarantees of the availability of computational resources, so an analysis of the fluctuation of the spot price is necessary to determine if spot instances should be utilized.

The spot instance price for various instance types in the three-month period from the 11\textsuperscript{th} of April to the 11\textsuperscript{th} July 2015 is shown in\footnote{In the case of \texttt{m4.4xlarge} and \texttt{m4.10xlarge} data was only available from the 11\textsuperscript{th} June.} Figure \ref{fig:FigSpot}. Evidently, the volatility of the spot price is proportional to number of CUs that an instance possesses. Therefore, while it would be difficult to rely on a \texttt{m4.10xlarge} spot instance, the spot price of the \texttt{m3.medium} spot instance is remarkably stable. Specifically, at no point in the three month period does the \texttt{m3.medium} spot price exceed \$$0.01$. Therefore, we can conclude that a significant reduction in cost can be achieved by using \texttt{m3.medium} spot instance with little effect on the reliability of the service, and with more flexibility than when using larger spot instances with more CUs.

\bibliographystyle{IEEEbib}
\bibliography{dithen_accepted_v24}

\begin{thebibliography}{10}

\bibitem{song2012optimal}
Y.~Song, M.~Zafer, and K.-W. Lee,
\newblock ``Optimal bidding in spot instance market,''
\newblock in {\em Proc. IEEE INFOCOM}, 2012, pp. 190--198.

\bibitem{zhang2014dynamic}
L.~Zhang, Z.~Li, and C.~Wu,
\newblock ``Dynamic resource provisioning in cloud computing: A randomized
  auction approach,''
\newblock in {\em Proc. of IEEE INFOCOM}, 2014.

\bibitem{nan2012optimal}
X.~Nan, Y.~He, and L.~Guan,
\newblock ``Optimal allocation of virtual machines for cloud-based multimedia
  applications,''
\newblock in {\em IEEE Int. Conf. Multimedia Signal Proc. (MMSP)}. IEEE, 2012,
  pp. 175--180.

\bibitem{hobfeld2012challenges}
T.~Hobfeld, R.~Schatz, M.~Varela, and C.~Timmerer,
\newblock ``Challenges of {QoE} management for cloud applications,''
\newblock {\em IEEE Comm. Mag.}, vol. 50, no. 4, pp. 28--36, 2012.

\bibitem{islam2012giving}
S.~Islam and J.-C. Gr{\'e}goire,
\newblock ``Giving users an edge: A flexible cloud model and its application
  for multimedia,''
\newblock {\em Fut. Gen. Comp. Syst.}, vol. 28, no. 6, pp. 823--832, 2012.

\bibitem{andreopoulos2008incremental}
Y.~Andreopoulos and M.~van~der Schaar,
\newblock ``Incremental refinement of computation for the discrete wavelet
  transform,''
\newblock {\em IEEE Trans. on Signal Process.}, vol. 56, no. 1, pp. 140--157,
  2008.

\bibitem{spiliotopoulos2001quantization}
V.~Spiliotopoulos et~al.,
\newblock ``Quantization effect on {VLSI} implementations for the 9/7 dwt
  filters,''
\newblock in {\em Proc. IEEE Int. Conf. on Acoust., Speech, and Signal
  Process., ICASSP'01}. IEEE, 2001, vol.~2, pp. 1197--1200.

\bibitem{andreopoulos2001local}
Y.~Andreopoulos et~al.,
\newblock ``A local wavelet transform implementation versus an optimal
  row-column algorithm for the 2d multilevel decomposition,''
\newblock in {\em Proc. IEEE Int. Conf. on Image Process., ICIP 2001}. IEEE,
  2001, vol.~3, pp. 330--333.

\bibitem{andreopoulos2002new}
Y~Andreopoulos et~al.,
\newblock ``A new method for complete-to-overcomplete discrete wavelet
  transforms,''
\newblock in {\em Proc. 14th IEEE Int. Conf. on Digital Signal Process., DSP
  2002}. IEEE, 2002, vol.~2, pp. 501--504.

\bibitem{kontorinis2009statistical}
N.~Kontorinis et~al.,
\newblock ``Statistical framework for video decoding complexity modeling and
  prediction,''
\newblock {\em IEEE Trans. on Circ. and Syst. for Video Technol.}, vol. 19, no.
  7, pp. 1000--1013, 2009.

\bibitem{masiyev2012cloud}
K.~Masiyev et~al.,
\newblock ``Cloud computing for business,''
\newblock in {\em IEEE Int. Conf. Appl. of Inf. and Comm. Tech. (AICT)}. IEEE,
  2012, pp. 1--4.

\bibitem{andreopoulos2000hybrid}
I~Andreopoulos et~al.,
\newblock ``A hybrid image compression algorithm based on fractal coding and
  wavelet transform,''
\newblock in {\em Proc. IEEE Int. Symp. Circuits and Systems, 2000 (ISCAS
  2000).} IEEE, 2000, vol.~3, pp. 37--40.

\bibitem{andreopoulos2003high}
Y.~Andreopoulos et~al.,
\newblock ``High-level cache modeling for {2-D} discrete wavelet transform
  implementations,''
\newblock {\em Journal of VLSI signal processing systems for signal, image and
  video technology}, vol. 34, no. 3, pp. 209--226, 2003.

\bibitem{Schwarzkopf2013Omega}
M.~Schwarzkopf et~al.,
\newblock ``Omega: Flexible, scalable schedulers for large compute clusters,''
\newblock in {\em Proceedings of the 8th ACM European Conference on Computer
  Systems}, 2013, EuroSys '13, pp. 351--364.

\bibitem{boutin2014apollo}
E.~Boutin et~al.,
\newblock ``Apollo: scalable and coordinated scheduling for cloud-scale
  computing,''
\newblock in {\em Proc. USENIX Symp. on Operating Systems Design and
  Implementation (OSDI)}, 2014.

\bibitem{ousterhout2013sparrow}
K.~Ousterhout et~al.,
\newblock ``Sparrow: distributed, low latency scheduling,''
\newblock in {\em Proceedings of the Twenty-Fourth ACM Symposium on Operating
  Systems Principles}, 2013, pp. 69--84.

\bibitem{Gandhi2012autoscale}
A.~Gandhi et~al.,
\newblock ``Autoscale: Dynamic, robust capacity management for multi-tier data
  centers,''
\newblock {\em ACM Trans. Comput. Syst.}, vol. 30, no. 4, pp. 14:1--14:26, Nov.
  2012.

\bibitem{Paya2015loadbalancing}
A.~Paya and D.~Marinescu,
\newblock ``Energy-aware load balancing and application scaling for the cloud
  ecosystem,''
\newblock {\em IEEE Trans. on Cloud Comp.}, vol. PP, no. 99, pp. 1--1, 2015.

\bibitem{ranjan2013mediawise}
R.~Ranjan, K.~Mitra, and D.~Georgakopoulos,
\newblock ``Mediawise cloud content orchestrator,''
\newblock {\em J. of Int. Serv. and Appl.}, vol. 4, no. 1, pp. 1--14, 2013.

\bibitem{jung2014estimation}
D.~Jung et~al.,
\newblock ``An estimation-based task load balancing scheduling in spot
  clouds,''
\newblock in {\em Network and Parallel Computing}, pp. 571--574. Springer,
  2014.

\bibitem{gulisano2012streamcloud}
V.~Gulisano et~al.,
\newblock ``Streamcloud: An elastic and scalable data streaming system,''
\newblock {\em IEEE Trans. Par. and Distr. Syst}, vol. 23, no. 12, pp.
  2351--2365, 2012.

\bibitem{Rodriguez2014deadlinescheduling}
M.A. Rodriguez and R.~Buyya,
\newblock ``Deadline based resource provisioning and scheduling algorithm for
  scientific workflows on clouds,''
\newblock {\em IEEE Trans. on Cloud Comp.}, vol. 2, no. 2, pp. 222--235, April
  2014.

\bibitem{Mao2011deadlinescheduling}
M.~Mao and M.~Humphrey,
\newblock ``Auto-scaling to minimize cost and meet application deadlines in
  cloud workflows,''
\newblock in {\em Proc. 2011 Int. Conf. High Perf. Comp., Netw., Stor. and
  Anal.}, 2011, pp. 49:1--49:12.

\bibitem{shorten2006positive}
R.~Shorten, F.~Wirth, and D.~Leith,
\newblock ``A positive systems model of {TCP}-like congestion control:
  asymptotic results,''
\newblock {\em IEEE/ACM Trans. Netw}, vol. 14, no. 3, pp. 616--629, 2006.

\bibitem{viola2004robust}
P.~Viola and M.~J Jones,
\newblock ``Robust real-time face detection,''
\newblock {\em Int. J. of Computer Vision}, vol. 57, no. 2, pp. 137--154, 2004.

\bibitem{imagemagick}
ImageMagick~Studio LLC,
\newblock ``Imagemagick,'' available online at: http://www.imagemagick.org/.

\bibitem{lowe2004sift}
D.~Lowe,
\newblock ``Distinctive image features from scale-invariant keypoints,''
\newblock {\em Int. J. Computer Vision}, vol. 60, no. 2, pp. 91--110, 2004.

\bibitem{Dean2008MapReduce}
J.~Dean and S.~Ghemawat,
\newblock ``{MapReduce}: Simplified data processing on large clusters,''
\newblock {\em Commun. ACM}, vol. 51, no. 1, pp. 107--113, Jan. 2008.

\bibitem{abbas2015vectors}
A.~Abbas, N.~Deligiannis, and Y.~Andreopoulos,
\newblock ``Vectors of locally aggregated centers for compact video
  representation,''
\newblock in {\em Proc. IEEE Int. Conf. Multimedia and Expo (ICME'15)}. IEEE,
  2015, pp. 1--6.

\bibitem{chadha2015region}
A.~Chadha and Y.~Andreopoulos,
\newblock ``Region-of-interest retrieval in large image datasets with {Voronoi
  VLAD},''
\newblock in {\em Computer Vision Systems}, pp. 218--227. Springer, 2015.

\bibitem{yang2004two}
J.~Yang et~al.,
\newblock ``Two-dimensional {PCA}: a new approach to appearance-based face
  representation and recognition,''
\newblock {\em IEEE Trans. Patt. Anal. and Machine Intel.}, vol. 26, no. 1, pp.
  131--137, 2004.

\bibitem{garcia2010study}
A.~Garcia, H.~Kalva, and B.~Furht,
\newblock ``A study of transcoding on cloud environments for video content
  delivery,''
\newblock in {\em Proc. 2010 ACM Multim. Workshop on Mob. Cloud Media Comput.}
  ACM, 2010, pp. 13--18.

\bibitem{Pouwelse2005tracker}
J.~Pouwelse et~al.,
\newblock ``The {Bittorrent P2P} file-sharing system: Measurements and
  analysis,''
\newblock in {\em Peer-to-Peer Systems IV}, vol. 3640 of {\em Lecture Notes in
  Computer Science}, pp. 205--216. Springer Berlin Heidelberg, 2005.

\bibitem{anderson2012optimal}
B.~D.~O. Anderson and J.~B. Moore,
\newblock {\em Optimal filtering},
\newblock Courier Corporation, 2012.

\bibitem{Margolies2014fairness}
R.~Margolies, A.~Sridharan, et~al.,
\newblock ``Exploiting mobility in proportional fair cellular scheduling:
  Measurements and algorithms,''
\newblock in {\em Proc. IEEE INFOCOM, 2014}. IEEE, 2014, pp. 1339--1347.

\bibitem{leutenegger2011brisk}
S.~Leutenegger, M.~Chli, and R.~Siegwart,
\newblock ``Optimal allocation of virtual machines for cloud-based multimedia
  applications,''
\newblock in {\em Proc. IEEE Int. Conf. Comp. Vis. (ICCV)}. IEEE, 2011, pp.
  2548--2555.

\bibitem{roy2011efficient}
N.~Roy, A.~Dubey, and A.~Gokhale,
\newblock ``Efficient autoscaling in the cloud using predictive models for
  workload forecasting,''
\newblock in {\em Proc. IEEE Int. Conf. on Cloud Com. (CLOUD)}. IEEE, 2011, pp.
  500--507.

\bibitem{debusschere2012hourly}
V.~Debusschere, S.~Bacha, et~al.,
\newblock ``Hourly server workload forecasting up to 168 hours ahead using
  seasonal {ARIMA} model,''
\newblock in {\em Proc. IEEE Int. Conf. on Industr. Technol.}, 2012.

\bibitem{Calheiros2014arima}
R.~Calheiros et~al.,
\newblock ``Workload prediction using arima model and its impact on cloud
  applications' {QoS},''
\newblock {\em IEEE Trans. on Cloud Comp.}, to appear.

\bibitem{tighe2014autoscale}
M.~Tighe and M.~Bauer,
\newblock ``Integrating cloud application autoscaling with dynamic {VM}
  allocation,''
\newblock in {\em IEEE Netw. Oper. and Manag. Symp. (NOMS)}, May 2014, pp.
  1--9.

\bibitem{krioukov2011napsac}
A.~Krioukov et~al.,
\newblock ``Napsac: Design and implementation of a power-proportional web
  cluster,''
\newblock {\em ACM SIGCOMM Comp. Comm. Rev.}, vol. 41, no. 1, pp. 102--108,
  2011.

\bibitem{chatfield2014return}
K.~Chatfield et~al.,
\newblock ``Return of the devil in the details: Delving deep into convolutional
  nets,''
\newblock {\em arXiv preprint arXiv:1405.3531}, 2014.

\bibitem{krizhevsky2012imagenet}
A.~Krizhevsky et~al.,
\newblock ``Imagenet classification with deep convolutional neural networks,''
\newblock in {\em Adv. in Neural Inf. Process. Syst. (NIPS'12)}, 2012, pp.
  1097--1105.

\bibitem{Hart1997Gutenberg}
M~Hart,
\newblock ``Project {Gutenberg},'' 1997,
\newblock http://www.gutenberg.net.

\bibitem{kouki2013scaling}
Y.~Kouki and T.~Ledoux,
\newblock ``{SCAling: SLA}-driven cloud auto-scaling,''
\newblock in {\em Proc. 28th Annual ACM Symp. on Applied Comp. (SAC'13)}. ACM,
  2013, pp. 411--414.

\bibitem{Choy2014AWSLargest}
S~Choy et~al.,
\newblock ``A hybrid edge-cloud architecture for reusing on-demand gaming
  latency,''
\newblock {\em Multim. Syst. J.}, vol. 20, no. 2, March 2014.

\end{thebibliography}


\vfill

\begin{IEEEbiography}[{\includegraphics[width=1in, height=1.0in,clip,keepaspectratio]{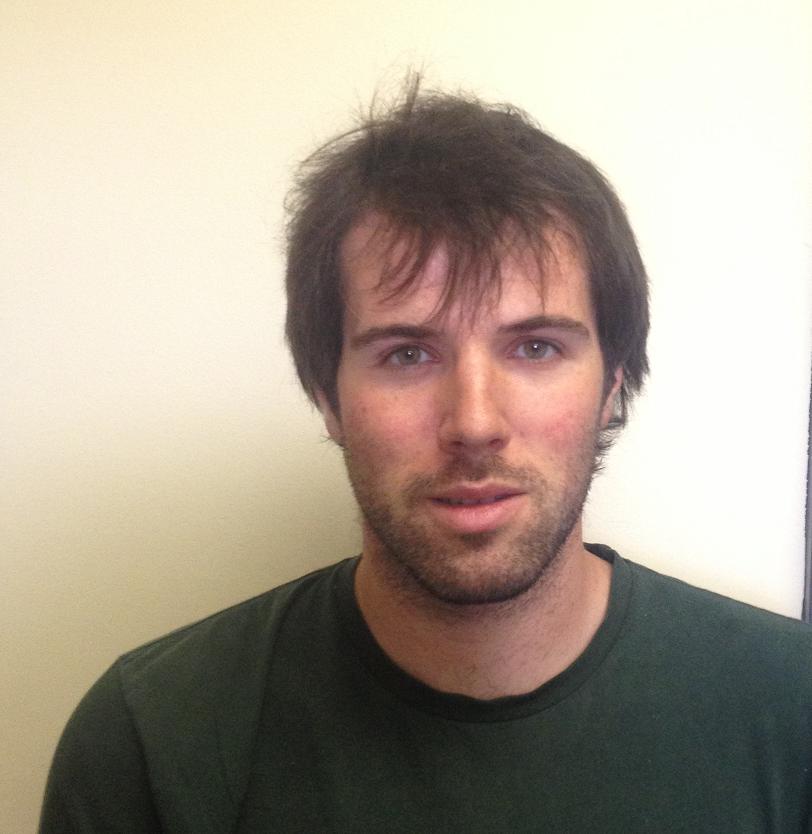}}]{Joseph Doyle} graduated from Trinity College Dublin in 2009 with a B.A.I., B.A. degree  in Computer and Electronic Engineering as a gold medalist. He was awarded a Ph.D in 2013 from Trinity College Dublin. He was a post-doctoral researcher in Trinity College Dublin and University College London from 2013 to 2014 and 2014 to 2016, respectively. He is a cofounder of Dithen Ltd. (London, U.K.) and is also Senior Lecturer at University of East London, London, U.K. His research interests include cloud computing, cognitive autoscaling, green computing,  and network optimization. 
\end{IEEEbiography}

\begin{IEEEbiography}[{\includegraphics[width=1in, height=1.0in,clip,keepaspectratio]{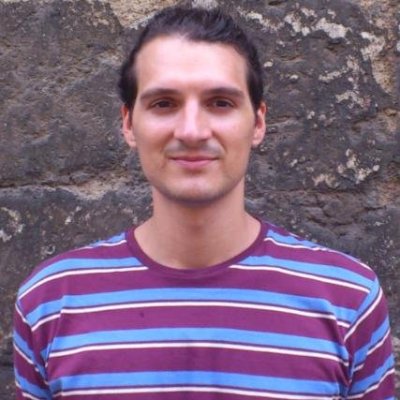}}]{Vasileios Giotsas} graduated with distinction from University College London (UCL) in 2008 with an MSc in Data Communications, Networks and Distributed Systems. He was awarded a Ph.D from University College London in 2014. He is a cofounder or Dithen Ltd. (London, U.K.) and is also a postdoctoral scientist at the UCSD Center for Applied Internet Data Analysis (CAIDA), La Jolla, CA. His research interests span the areas of distributed systems, cloud computing, routing protocols, Internet measurements, and Internet economics.
\end{IEEEbiography}

\begin{IEEEbiography}[{\includegraphics[width=1in, height=1.0in,clip,keepaspectratio]{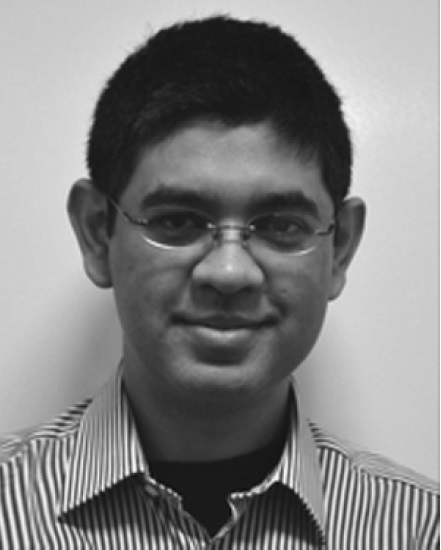}}]{Mohammad Ashraful Anam} obtained the PhD\ in Electronic Engineering from University College London (Lombardi Prize for the Best PhD\ thesis in Electronic Engineering)\ and is a cofounder of Dithen Ltd. (London, U.K.), as well as post-doctoral research associate in the Department of Electronic and Electrical Engineering, University College London, London, UK. His research interests are in error tolerant computing, and reliable cloud computing.  \end{IEEEbiography}

\begin{IEEEbiography}[{\includegraphics[width=1in, height=1.0in,clip,keepaspectratio]{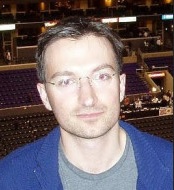}}]{Yiannis Andreopoulos}(M'00--SM'14) obtained the Electrical Engineering Diploma and an MSc from the University of Patras, Greece, and the PhD\ in Applied Sciences from the Vrije Universiteit Brussel, Belgium. He is a cofounder of Dithen Ltd. (London, U.K.), as well as Reader (Assoc. Professor) in Data and Signal Processing Systems in the Electronic and Electrical Engineering Department of University College London, London, U.K. His research interests are in error-tolerant computing and multimedia systems.  \end{IEEEbiography}

\vfill
\enlargethispage{-5in}

\end{document}